\documentclass{aa}
\usepackage{graphicx} 
\usepackage{txfonts}

\def\sec{\ifmmode {}^{\prime\prime}\else ${}^{\prime\prime}$\fi~}

\def\magdot{\ifmmode {}^{\rm m}\!\!\!.\, \else ${}^{\rm m}\!\!\!.\,$\fi}
\def\daydot{\ifmmode {}^{\rm d}\!\!\!.\, \else ${}^{\rm d}\!\!\!.\,$\fi}
\def\asec{\ifmmode ^{\prime\prime}\else$^{\prime\prime}$\fi}

\begin{document}

\authorrunning{Matute, La Franca, Pozzi et al.}
\title{Active Galactic Nuclei in the mid-IR}
\subtitle{Evolution and Contribution to the Cosmic Infrared Background
\thanks{Data presented here are based on observations by the Infrared Space
Observatory ({\it ISO}). {\it ISO} is an ESA project with instruments funded by ESA
member states (especially the PI countries: France, Germany, the Netherlands and the
United Kingdom) and with the participation of ISAS and NASA.}
}

\author{I.\,Matute \inst{1,2}, F.\,La Franca \inst{2},
F.\,Pozzi \inst{3,4}, C.\,Gruppioni \inst{4}, C.\,Lari \inst{5}, 
G.\,Zamorani \inst{4}}

\institute{
Max-Planck Institut f\"ur extraterrestrische Physik (MPE), 
Giessenbachstra$\beta$e, Postfach 1312,
D-85741 Garching, Germany \\
\email{matute@mpe.mpg.de}
\and
Dipartimento di Fisica , Universit\`a  degli Studi
``Roma Tre'', Via della Vasca Navale 84, I-00146 Roma, Italy \\
\email{lafranca@fis.uniroma3.it}
\and
Dipartimento di Astronomia, Universit\`a di Bologna, 
         Via Ranzani 1, I-40127 Bologna, Italy
\and
INAF, Osservatorio Astronomico di Bologna, Via Ranzani 1, 
                I-40127 Bologna, Italy
\and
INAF, Istituto di Radioastronomia (IRA), Via Gobetti 101, 
                I-40129 Bologna, Italy
}

\date{Received June 28, 2005; accepted December 30, 2005}

  \abstract
    {}
   {We study the evolution of the luminosity function (LF) of type-1 and type-2
    Active Galactic Nuclei (AGN) in the mid-infrared, and then derive the
    contribution of the AGN to the Cosmic InfraRed Background (CIRB) and the
    expected source counts to be observed by $S\!pitzer$ at 24\,$\mu$m.}
  {We used a sample of type-1 and type-2 AGN selected at 15$\,\mu$m
    ({\it ISO}) and 12$\,\mu$m ({\it IRAS}), and classified on the
    basis of their optical spectra. Local spectral templates of type-1 and
    type-2 AGN have been used to derive the intrinsic 15$\,\mu$m
    luminosities.  We adopted an evolving smooth two-power law
    shape of the LF, whose parameters have been derived using an
    un-binned maximum likelihood method.}
  {We find that the LF of type-1 AGN is compatible
    with a pure luminosity evolution ($L(z)=L(0)(1+z)^{k_L}$) model
    where $k_L\sim$2.9.  A small flattening of the faint
    ($L_{15}<L^{*}_{15}$) slope of the LF with
    increasing redshift is favoured by the data.  A similar
    evolutionary scenario is found for the type-2 population with a
    rate $k_L$ ranging from $\sim$1.8 to 2.6, depending significantly
    on the adopted mid-infrared spectral energy distribution.  Also
    for type-2 AGN a flattening of the LF with
    increasing redshift is suggested by the data, possibly caused by
    the loss of a fraction of type-2 AGN hidden within the optically
    classified starburst and normal galaxies. The type-1 AGN
    contribution to the CIRB at 15$\,\mu$m is
    (4.2--12.1)$\times\,10^{-11}\;\mathrm{W m^{-2} sr^{-1}}$, while
    the type-2 AGN contribution is
    (5.5--11.0)$\times\,10^{-11}\;\mathrm{W\,m^{-2}\,sr^{-1}}$.  We
    expect that $S\!pitzer$ will observe, down to a flux limit of $S_{24\,\mu
      \mathrm{m}}$=0.01\,mJy,  a density of $\sim$1200 deg$^{-2}$ type-1 and
    $\sim$1000 deg$^{-2}$ type-2 optically classified AGN.}
   {AGN evolve in the mid-infrared  with a rate
   similar to the ones found in the optical and X--rays bands. The derived total
   contribution of the AGN to the CIRB (4-10\%) and $S\!pitzer$ counts should be
   considered as lower limits, because of a possible loss of type-2 sources
   caused by the optical classification.}


\titlerunning{The Evolution of AGN in the mid-IR}
\maketitle

\section{Introduction}

Active Galactic Nucleus (AGN hereafter) are expected to have played 
an important role in the formation and evolution of the galaxies in the
Universe. An example is the observation of the correlation between the mass of
the central black holes ($M_{BH}$) and the mass of the bulges (Magorrian et al.
1998) or the velocity dispersion of gas and stars in the bulges of galaxies
($M_{BH}$--$\sigma$ relation; Ferrarese \& Merritt 2000; Tremaine et al. 2002).
Thus, the measure of the history of the density of AGN as a function of the
luminosity (the luminosity function, LF hereafter) can provide fundamental clues
to explain the present day universe (e.g. Balland et al.  2003; Menci et al.
2004; Granato et al. 2004; Di Matteo et al. 2005).

The LF will not only provide informations on the demographics of AGN, but will also 
place constraints on the physical model of AGN, the origin and accretion history 
into supermassive black hole and the formation of structures in the early universe. 
Moreover, the measure of the AGN LF over the whole spectral range will provide 
information on the relative importance of accretion power in the overall energy
budget of the universe.

AGN have been historically classified into two groups (type-1 and type-2), 
depending on the presence or absence in their optical spectra of
broad emission lines (FWHM$>$1200 $km\,s^{-1}$). The unified model
for AGN assumes that the two types of sources are
intrinsically the same (Antonucci et al. 1993), and that the observed
differences in the optical spectra are explained by an orientation
effect. The presence of obscuring material ({\it torus} or {\it warped
disc}) around the central energy source and its orientation respect to 
the observer could prevent a direct view of the central region around 
the nuclei responsible for the broad line emission observed in the optical 
band. AGN are then classified in two groups: the un-obscured (type-1) and 
obscured (type-2) sources depending  whether the line-of-sight intersects or
not the obscuring material. However, nowdays it seems clear that the 
orientation-dependent model is probably just a 0th-order approximation to 
the true nature of AGN (e.g. Barger et al. 2005; La Franca et al. 2005). 

The major efforts to understand the evolution of AGN have been
historically concentrated at optical wavelengths and in the last years 
have produced the largest compilations of AGN from the Two Degree Field
(2dF; Boyle et al. 2000; Croom et al. 2004) and the Sloan Digital Sky Survey
(SDSS; York et al. 2000; Hao et al. 2005). However the optical bands have
proved to be very inefficient in the selection of obscured
sources and the assembly of unbiased samples of type-2 objects
is difficult.  In the last years many new windows have been opened
at various wavelength bands for the observation of high redshift galaxies
and AGN: sub-mm ({\it SCUBA}), Infrared ({\it IRAS}, {\it ISO} and {\it
Spitzer}) and the X--ray ({\it ROSAT}, {\it XMM} and {\it Chandra}).
Thanks to their smaller dependence on dust obscuration if compared to 
the optical surveys, the Infrared and the hard X--ray wavelengths have 
proved to be much more efficient in the detection of type-2 sources.

X--ray observations have been so far consistent with population
synthesis models based on the 0th-order unified AGN scheme (e.g.
Comastri et al. 1995) or its modifications (Pompilio et al. 2000;
Gilli et al. 2001; Ueda et al. 2003; La Franca et al. 2005).  The
hard spectrum of the X--ray background is explained by a mixture of
absorbed and unabsorbed AGN, evolving with cosmic time. According to
these models, most AGN spectra should be heavily obscured as the light
produced by accretion is absorbed by gas and dust.

For this reason an important mid-infrared (4-40\,$\mu$m) thermal
emission is expected from reprocessed radiation by gas and
dust grains directly heated by the central black hole (e.g. Granato et
al. 1997; Oliva et al. 1999; Nenkova et al. 2002). AGN could be in
this case important contributors to the Cosmic Infrared Background
(CIRB). Indeed, {\it IRAS} has observed strong mid-IR emission in all local
($z<0.1$) AGN (Miley et al. 1985; de Grijp et al. 1985; Neugebauer et al. 1986; 
Sanders et al. 1989). Complete and largely unbiased samples of AGN have been 
produced at 12$\,\mu$m (Rush, Malkan \& Spinoglio 1993, RMS hereafter) and
25$\,\mu$m (Shupe et al. 1998) based on {\it IRAS} observations.  These
samples provide an important understanding of the infrared emission
from local galaxies and a firm base to compare the properties of
galaxies in the local universe with the high redshift populations uncovered
by {\it ISO} ($z\leq 1$) and most recently by {\it Spitzer} ($z\leq 2$).

Statistical studies of AGN in the mid-IR have been based on large but
local samples from the {\it IRAS} observations (e.g. RMS, Shupe et al.
1998) and deep but small {\it ISO} fields in the {\it Hubble Deep
Field North} (HDFN; Aussel et al. 1999a, 1999b), {\it  South} (HDFS; 
Oliver et al. 2002) and the {\it Canada-France Redshift Survey} (CFRS; 
Flores et al. 1999) that provided only a handful of objects.

Therefore, due to the small number of high redshift objects available, the
shape and evolution of the mid-IR LF of these objects at high redshift was
largely unknown. Only recently, large and highly complete 15$\,\mu$m
spectroscopic samples, based on the {\it European Large Area ISO
Survey} (ELAIS hereafter; Oliver et al. 2000), have been released
(Rowan-Robinson et al.  2004; La Franca et al. 2004; FLF04 hereafter).
Based on the analysis of the {\it ISO} observations in the ELAIS-S1
field, Matute et al. (2002, M02 hereafter) derived the first estimate
of the type-1 LF in the mid-IR (15\,$\mu$m)  finding an evolution similar
to the ones observed in the optical (Croom et al. 2004) and X--rays 
(e.g. Hasinger et al. 2005; La Franca et al. 2005). The evolution of the LF
for normal and starburst galaxies detected at 15$\mu$m in the southern ELAIS
fields (S1 \& S2) has been derived and discussed by Pozzi et al. (2004).

In this paper we investigate the separate evolution of type-1 and
type-2 AGN in the mid-IR and their contribution to the CIRB using: $a$)
the local IR population uncovered by {\it IRAS}, $b$) deep observations
from {\it ISO} fields and, $c$) the most recent spectroscopic
identifications of 15$\mu$m sources in the southern fields of the
ELAIS Survey (Pozzi et al. 2003; FLF04)\footnote{Data and
related papers about the ELAIS southern survey are available at:
{\sf http://www.fis.uniroma3.it/$\sim$ELAIS\_S}}.  The AGN sample
used in our work is described in section 2. Section 3 introduces the
general method used to derive the parameters of the luminosity
function and presents the results. In section 4 we discuss our results
and compare them with the LFs derived in the mid-IR by previous
analysis of mid-IR observations. The contribution to the CIRB and the
predicted counts in the mid-IR {\it Spitzer} band at 24\,$\mu$m are
computed and discussed in sections 5 and 6. Section 7 summarises the
main results.

A cosmology with H$_\mathrm{o}=75$ km s$^{-1}$ Mpc$^{-1}$ and 
$\Omega_{\mathrm \lambda}=0.7$ , $\Omega_{\mathrm m}=0.3$ was adopted 
for the work presented in this paper.

\section{The sample and source classification}

The mid-IR selected AGN sample used in this work has been extracted
from:
($i$) the 15$\, \mu$m ELAIS fields S1 and S2 
        (Lari et al.  2001; Pozzi et al. 2003), 
($ii$) the 15$\, \mu$m deep {\it ISO} surveys in the HDFN 
(Aussel et al. 1999a, 1999b), HDFS (Oliver et al. 2002) and CFRS (Flores et al. 1999), 
($iii$) the local {\it IRAS} 12$\, \mu$m sample of RMS. 

The 15$\, \mu$m catalogue in the ELAIS S1 field has been released
by Lari et al. (2001). It covers an area of $\sim$4 deg$^2$ centred at
$\alpha(2000)=00^h 34^m 44.4^s$, $\delta(2000)=-43^o 28' 12''$ and includes 462
mid-IR sources down to a flux limit of 0.5 mJy. Mid-IR source counts based on this 
catalogue have been presented and discussed by Gruppioni et al. (2002).

\begin{figure}[!t]
\label{fig:rmagvsz}
\centering
\resizebox{\hsize}{!}{\includegraphics{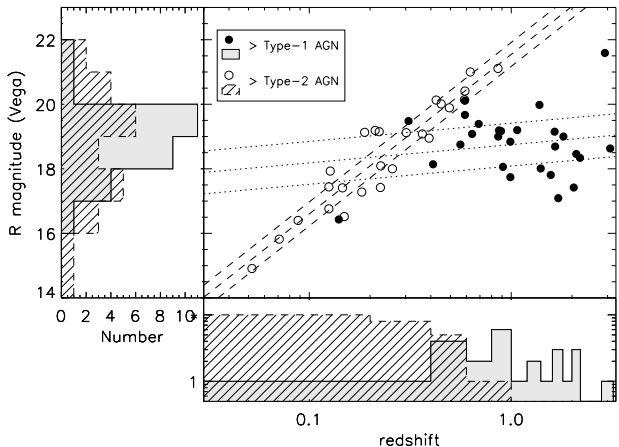}}
\caption{\scriptsize Distribution of ELAIS-S1 type-1 and -2 AGN in the 
$z-R$-magnitude plane. Best fit linear solution ($\pm$1$\sigma$) in the
log($z$)-R space are shown as {\it dotted} lines for type-1 AGN and as {\it
dashed} lines for type-2 AGN. The $R$-magnitude and redshift distributions are
shown in the {\it left} and {\it bottom} panels respectively.}
\end{figure}

The analysis presented here is restricted to the more reliable subsample of 
406 sources as described by FLF04.  About 80$\%$ of these sources have been
optically identified on CCD exposures down to {\it R}$\sim$23, while spectral
classification has been obtained for 90$\%$ of the optically identified sample.
As discussed by FLF04, due to a different mid-IR flux limit coverage of the
ELAIS-S1 field, the total area has been divided into two regions: the central
and deepest region of S1 (S1-5) reaching mid-IR fluxes ($S_{15}$ hereafter) of
0.5 mJy, and the outer region (S1-rest) with a 0.9 mJy flux limit. The S1-5 area
is spectroscopically complete at the 97$\%$ level down to $R$=21.6, while
S1-rest completeness reaches the 98$\%$ level down to $R$=20.5. In total 116 
sources (29$\%$  of the total mid-IR sample) do not have a spectroscopic
identification due to incompleteness of the follow-up or to the lack of optical 
counterpart brighter than $R$=23. A detailed description of the optical 
identification, spectroscopic classification, size and completeness function of
the different areas used in the ELAIS-S1 sample are presented and discussed by
FLF04.

The S2 field is a smaller and deeper area centred at $\alpha(2000)=05^h 02^m 24.5^s$,
$\delta(2000)=-30^o 36' 00''$  covering $\sim$ 0.12 deg$^2$. This region
includes 43 sources with {\it S/N}$\,>\,$5 down to $S_{15}$=0.4 mJy
(Pozzi et al. 2003; Rowan-Robinson et al. 2004). Photometry in the whole field 
is available in the {\it U}, {\it B}, {\it I} and {\it K$^\prime$} bands down 
to 21.0, 24.5, 22.0 and 18.75 respectively (Vega magnitudes). 
39 infrared sources have a counterpart in the {\it I} band. 22 of them have a 
spectroscopic identification, while 8 sources are classified as stars since they
are associated with bright ({\it I}$<$14) point-like sources in the optical
catalogue. To avoid large incompleteness due to the optical follow-up, only
sources with optical counterparts brighter than {\it I}$=$20.6 were selected. Down
to this optical limit, the sample considered in S2 is then complete at the 95\%
spectroscopic level for sources with $S_{15}>$0.7 mJy. 

Classification for AGN dominated sources in the ELAIS fields was based on 
their optical spectral signatures.  Sources showing broad emission line 
profiles (rest frame FWHM$>$1200 $km s^{-1}$) were classified as type-1. 
Type-2 sources were selected following classic diagnostic diagrams (e.g. Tresse 
et al. 1996; Osterbrock 1989; Veilleux \& Osterbrock 1987) that included one or 
more of the following line ratios: [NII]/H$\alpha$, SII/(H$\alpha$+[NII]),
OI/H$\alpha$, [OIII]/H$\beta$ and [OII]/H$\beta$ when available, 
depending on the redshift of the source ({\it e.g.} log([OIII]/H$\beta$)$>$0.5 
and log([NII]/H$\alpha$)$>$-0.4).

In total, the southern ELAIS (S1+S2) AGN sample includes 27 (25+2) type-1 and
25 (23+2) type-2 AGN representing $\sim$16$\%$ (52/320) of the identified
infrared sources and $\sim$24$\%$ (52/221) of the extragalactic population.
Their distribution in the redshift--$R$-magnitude space is shown in Fig. 1. The
plotted lines represent a linear fit ($\pm1\sigma$)  to the data ({\it dotted}
for type-1 AGN and {\it dashed} for type-2). Type-1 sources are detected up to
$z\!\sim$3 and show a quasi redshift-independent  and narrow optical magnitude
distribution ($R=18.76\pm1.01$). On the other hand, type-2 sources show a larger
spread of optical magnitudes ($R=18.45\pm1.62$) and a rather steep dependence on
the redshift up to $z\!\sim$1. 

Constraints on the fainter population are provided by deep 15$\,\mu$m
observations in the HDFN (Aussel et al. 1999a, 1999b), HDFS (Oliver et al. 2002) 
and in the 1452+52 field of the CFRS (Flores et al. 1999). These fields have a
flux limit  about an order of magnitude deeper ($S_{15}\simeq0.1$ mJy)
than the ELAIS-S fields but with sky coverages 100 to 500 times
smaller. The multiwavelength follow-up in these fields has identified
a small fraction of sources as AGN (both type-1 and type-2).

In the HDFN, 41 15$\,\mu$m sources  have been detected  over an area of 21.5
arcmin$^2$.  The sample is complete down to 0.1 mJy (Aussel et al. 1999a). 4 of
these sources,  with fluxes between 0.1 and 0.45 mJy, are found to be AGN  by
Alexander et al. (2002), equally divided between the two classes (Alexander et
al., private communication). In the HDFS we use the AGN identified by
Franceschini et al. (2003)  from a sample of 59 ISOCAM sources with
$S_{15}\!\ge\!0.1$ mJy  selected over an area of 19.6 arcmin$^2$.  They have
detected 2 secure AGN, one type-1 (S$_{15}=0.288$ mJy) and one type-2
(S$_{15}=0.518$ mJy). The observations in the CFRS 1452+52 field cover an area
of $\sim$100 arcmin$^2$. 41 sources with $S_{15}\ge0.35$ mJy and with S/N$\ge$4
have been selected  (Flores et al. 1999).  The spectroscopic follow-up has
identified one type-1 ($S_{15}\!=\!0.459$ mJy) and one type-2
($S_{15}\!=\!1.653$ mJy) source. In the above mentioned fields, a fraction of
the type-2 sources  were not classified according to their optical spectra but
on the basis of the shape of their multiwavelength spectral energy distribution
(SED). For this reason only the selected type-1 population has been used in our
analysis.

Finally, we have combined the ISOCAM data with the local {\it IRAS}
sample of AGN from RMS. The RMS Catalogue is a high galactic latitude
($|b|\!\ge\! 25^\mathrm{o}$), colour selected ($S_{60 \mu
  \mathrm{m}}$/$S_{12 \mu \mathrm{m}}$) and flux limited ($S_{12 \mu
  \mathrm{m}}\!>$0.22 Jy) sample extracted from the {\it IRAS}
12$\,\mu$m {\it Faint Source Catalog, Version 2} (Moshir et al. 1991).
To avoid completeness uncertainties in the mid-IR sample, only sources
with $S_{12\mu \mathrm{m}}\!\ge$300 mJy were selected (see RMS).
In total 41 type-1 and 50 type-2 sources are found within this flux range. 
The photometric and spectroscopic follow-up of the optical counterparts is 
considered to be $\sim$100\% complete for these sources and has been used in 
our analysis as representative of the mid-IR selected population of AGN in 
the local universe.

\begin{table}
\label{mean_z_L}
  \caption{Mean redshift and Luminosity ($\pm 1\sigma$) of the mid-IR samples}
  \centering
  \begin{tabular}{ccccc}
    \hline\hline
     Sample & \multicolumn{2}{c}{\bf  IRAS } & \multicolumn{2}{c}{\bf  ISO  } \\
     & $<z>$ & $<\mathrm{log}\, L_{15}>^{(a)}$ & $<z>$ & $<\mathrm{log}\, L_{15}>$ \\
    \hline
    Type-1 & 0.03$\pm$0.02 & 43.87$\pm$0.87 & 1.23$\pm$0.71 & 45.22$\pm$0.78 \\
    Type-2 & 0.02$\pm$0.01 & 43.66$\pm$0.68 & 0.28$\pm$0.20 & 43.90$\pm$0.69 \\

    \hline
  \end{tabular}
        \begin{itemize}
        \item[$^{(a)}$] \scriptsize{12\,$\mu$m luminosities from RMS have been 
        converted to 15\,$\mu$m using the adopted SED. In particular the $Circinus$ 
        SED was adopted for type-2 AGN (see \S 3.1).}
        \end{itemize}
\end{table}

\begin{figure}[!ht]
\centering
\resizebox{\hsize}{!}{\includegraphics{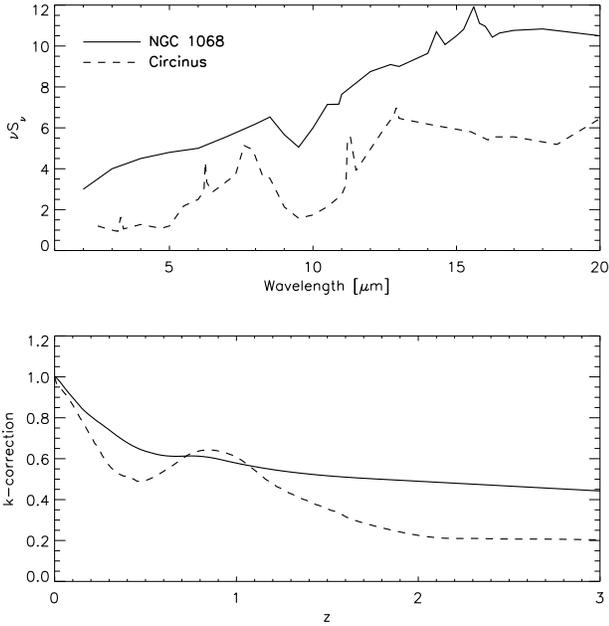}}
\caption{\scriptsize Adopted mid-IR spectral energy distributions (SEDs) 
for type-2 sources ({\it top} panel) and  the corresponding 
$k$-corrections as a function of redshift computed for the ISOCAM-LW3 
filter ({\it bottom} panel).}
\label{fig:SEDs}
\end{figure}

\section{The Evolution of AGN}

\subsection{The Method}

In order to compare local sources with a higher redshift selected population and
to probe changes with cosmic time it is necessary to compare their intrinsic
physical properties. Our sample  has been selected in the mid-IR and identified
in the optical band, therefore rest-frame luminosities, in the form of
$\nu$L$_\nu$, at 15$\,\mu$m ($L_{15}$) and in the $R$-band ($L_{R}$) were
computed using well known SEDs for type-1 and 2 sources.

The compilation of Elvis et al. (1994) of 47 QSO, in the range of 1--20$\,\mu$m, 
was used as representative of the mid-IR emission for the type-1 population. 
This composite SED agrees very well with spectroscopic observations of type-1
sources performed by {\it ISO} in the mid-IR (Clavel et al. 2000; Spoon et al.
2002), showing a strong power-like continuum, very weak or no PAH emission bands
and no evidence of the 10$\,\mu$m silicate absorption feature.  Optical
luminosities for type-1 AGN were computed using the $R$-band $k$-correction by
Natali et al. (1998), which takes into account the large effect of broad lines
entering and leaving the passband. 

Unlike type-1 sources, the mid-IR SED of type-2 objects varies greatly. They
range from starburst-like galaxies SEDs, as in the case of Circinus (Sturm et
al. 2000), showing a weak continuum, prominent emission from PAH molecules and a
deep silicate absorption feature at $\sim$10$\,\mu$m, to more power-like SEDs
dominated by hot dust directly heated by the active nucleus, as in NGC 1068
(Sturm et al. 2000).  As a consequence, the mid-IR SED of NGC 1068 and Circinus
galaxies (Fig. 2) were used as representative of two extreme cases of obscured
AGN in the mid-IR. In order to derive the optical $k$-correction for  type-2
sources, an average spectrum was produced in the optical band (3500-7000\AA)
using our best type-2 spectra (Fig. 5 in FLF04).

In Figure\,3 we show the distributions in the $L_{15}-z$ plane of the total 
sample used in our analysis, while Table\,1 summarises the mean redshifts and
mid-IR luminosities, and corresponding 1$\sigma$ dispersion, measured for 
the local ({\it IRAS}) and high redshift ({\it ISO}) sources. 

\begin{figure}[!ht]
\centering
\resizebox{\hsize}{!}{\includegraphics{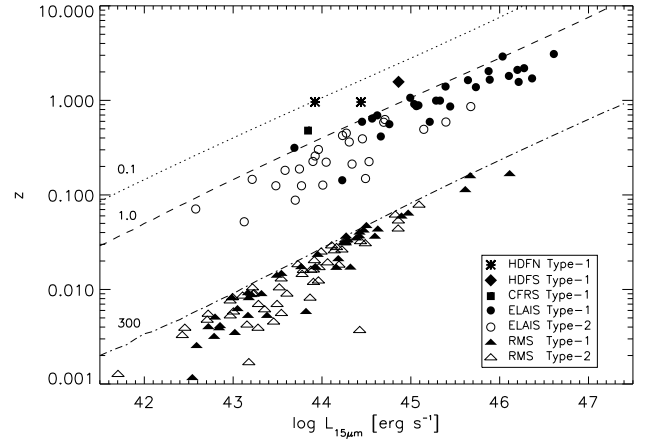}}
\caption{\scriptsize Distribution in the  $L_{15}- z$ space of the total sample 
of mid-IR type-1 and 2 AGN. The mid-IR luminosity for type-2 sources has been
computed applying the Circinus $k$-correction (see text, $\S$3.1). The dotted,
dashed and dot-dashed lines represent the flux limits at 0.1, 1.0 and 300 mJy
respectively.}
\label{fig:L_z}
\end{figure}

The adopted shape of the luminosity function (LF) is a smooth two-power law 
of the form:

\begin{center}
\begin{equation}
\frac{d\Phi({L}_{\rm 15},z)}{d{\rm log}{L}_{\rm 15}} =
\frac{\Phi^{*}}{(L_{\rm 15}/{{L}^{*}_{\rm
15}})^{\alpha(z)} + (L_{\rm 15}/{{L}^{*}_{\rm15}})^{\beta}}
\end{equation}
\end{center}

where we have introduced a dependence on $z$ in order to allow a change 
in the slope ($\alpha$) of the faint end of the luminosity function. 
We have assumed a luminosity evolution of the form: 
\begin{center}
\begin{equation} \label{eqn:LF}
L_{\rm 15}(z)= \left\{ \begin{array}{ll}
L_{\rm 15}(0)(1+z)^{k_{L}} & \mbox{for  $z \le z_{cut}$}\\
L_{\rm 15}(z_{cut}) & \mbox{for  $z > z_{cut}$}
\end{array} \right.
\end{equation}
\end{center}
where $z_{cut}$ is the redshift cut beyond which the luminosity function 
is assumed to remain constant. 

The parameters for the luminosity function and the evolution have been
derived using an un-binned, maximum likelihood method (Marshall et al.
1983), as described by Matute et al. (2002). The spectro-photometric
follow-up of the ELAIS southern fields was not deep enough to provide
an identification and a classification for all the sources detected in
the mid-IR sample. A factor depending on redshift and mid-IR luminosity was 
therefore introduced in order to take into account the fraction of objects not
identified, due to the lack of an optical counterpart or to incompleteness of
the spectroscopic follow-up.  We will refer to this factor as the
$optical-completeness$ factor, $\mathbf{\Theta}(z,L)$. It represents
the probability that a source with a given redshift and mid-IR
luminosity $L_{\rm 15}$ has an apparent R-band magnitude within the
spectroscopic limits of the samples, and was derived taking into
account both the average $L_{15}$/$L_R$ ratio of the sources {\it and its
observed natural spread}. For any given mid-IR luminosity $L_{15}$, 
this probability has been computed assuming a gaussian distribution of 
log\,($L_{15}/L_R$) centred at its mean value and with a sigma equal to the
observed natural spread of the $L_{15}$/$L_R$ relation (see next sections for
discussions on the adopted mean values and spreads of log\,($L_{15}/L_R$))
\footnote{A similar approach was adopted in the previous estimates of the
MIR type-1 AGN LF by Matute et al. (2002), and of the X-ray AGN LF by La Franca
et al.\  (2002, 2005)}. Then, the function to be minimised can be written as:
\begin{equation}
\mathrm{S}=
-2\sum_{i=1}^{\mathrm{N}} \ln[\Phi(z_{i},\!L_{i})]  + 2\!\!\int\!\!\!\!\!\int 
\!\Phi(z,\!L) \Omega(z,\!L)  \mathbf{\Theta}(z,\!L)  \frac {\textstyle dV}
{\textstyle dz}dzdL
\end{equation}
where $\Omega(z,\!L)$ is the effective area covered by each subsample at
15$\,\mu$m, {\it L}=$L_{15}$ and the index {\it i} runs over all the objects of
a given class. For the ELAIS-S1 sources the area coverage is presented as a
function of flux by FLF04. Area coverages for the deeper ISO fields (HDF-N, -S
and CFRS) are described in the previous section. The area sampled by RMS is
considered equal to 0 for $S_{12}<300$ mJy and equal to 22191.80 deg$^2$  for
$S_{12}\ge 300$ mJy (see RMS for details).  The representative SEDs were used to
convert the RMS 12$\,\mu$m fluxes into 15$\,\mu$m fluxes.

Confidence regions for each parameter have been obtained by computing $\Delta
S(=\Delta \chi^2)$ at a number of values around the best parameter ($S_{min}$), 
while allowing the other parameters to float (see Lampton, Margon
\& Bowyer 1976). A 68$\%$ confidence region  for any single parameter 
corresponds to  $\Delta S\!=\!1$. The goodness of the fit has been 
verified using the bidimensional Kolmogorov-Smirnov test (2D-KS hereafter) in
the $L_{15}$--$z$ space (see Peacock 1983; Fasano \& Franceschini 1987). The
normalisation factor $\Phi^{*}$ is determined in such a way to reproduce the
observed total number of sources ({\it ISO} + {\it IRAS}).

\subsection{The Evolution of type-1 AGN}

For type-1 sources, the optical completeness factor, $\Theta(z,L_{15})$, has
been estimated using the mean value log($L_{15}$/$L_R$)=0.23 with a 1$\sigma$
dispersion of 0.25 (from the sample of Elvis et al.\ 1994).  These values are in
rough agreement with the derived value for the {\it IRAS} local sub-sample of
type-1 AGN from Spinoglio et al.  (1995; 0.05 with a 1$\sigma$ spread of 0.13).
This ratio and its spread shows no significant dependence on $L_{15}$ over more
than 4 decades of mid-IR luminosity and we assumed them to be constant with
redshift.

In order to find the best fit solutions to the LF we have considered:
\begin{itemize}
\item [-] 2 different behaviours for the faint slope of the LF:  a) a
  fixed value ($\alpha=\alpha(0)$) and, b) a dependence of $\alpha$
  on $z$ of the form: $\alpha(z)=\alpha_1 \,
  exp({-\alpha_2}*{z})+\alpha_3$. A justification for this form is
  given in the following paragraphs.
\item [-] A fixed value for $z_{cut}$ equal to 2.0 in our
  parameterisation of the LF. Our relatively small number of sources, especially
  at high $z$, does not allow to sample correctly the existence of this
  redshift cutoff. We have therefore assumed a behaviour in the mid-IR similar
  to what is already known for the evolution of AGN in the optical (Boyle et al.
  2000, Croom et al. 2004) and X--ray (Miyaji et al. 2000, 2001; La Franca
  et al. 2005; Hasinger et al. 2005), suggesting values for $z_{cut}$
  in the range 1.7-2.5.
\item [-] A universe volume between $z$=0 and $z$=4,  while the
        space density was integrated in the luminosity interval log$L_{15}$=[42,47].
\end{itemize}

\begin{table*}[!t]
\caption{Parameter values of the fit of the Luminosity Function}
     \begin{center}
        \begin{tabular}{ccccccccccrc}
        \hline

& Model & $\alpha _1$ & $\alpha _2$ & $\alpha _3$ & $\beta$ & 
${\mathrm{log} \, L_{15}^*}$ $^a$ & $k_{L}$ & $z_{cut}$ & 
${\mathrm{log \,\Phi^{*}}}$ $^b$ & $\mathrm{CIRB}\,(\%)^{c}$ & $P_{2DKS}$ \\
\hline
 & \tiny{(I)} & \tiny{(II)} & \tiny{(III)} & \tiny{(IV)} & \tiny{(V)} & 
\tiny{(VI)} & \tiny{(VII)} & \tiny{(VIII)} & \tiny{(IX)} & \tiny{(X) \hspace{0.6cm}} &
 \tiny{(XI)} \\
\hline \hline

\multicolumn{12}{c}{\bf Type-1 AGN} \\
\hline
 & A &
$0.00\,$({\scriptsize{fixed}}) & $0.00\,$({\scriptsize{fixed}}) & $1.08_{-0.11}^{+0.09}$ &
$2.29_{-0.16}^{+0.36}$ & $44.45_{-0.08}^{+0.08}$ & ${2.78}_{-0.18}^{+0.23}$ & 
$2.00$ & $-5.78\mathbf{\pm0.03}$ & $12.1\mathbf{^{+7.7}_{-2.5}} \,(\mathbf{4.5^{+2.9}_{-1.0}})$ & $0.11$ \\
 & B & 
$0.69_{-0.08}^{+0.07}$ &$5.92_{-0.12}^{+0.17}$ & $0.64_{-0.11}^{+0.07}$ &
$2.76_{-0.26}^{+0.21}$ &$44.78_{-0.06}^{+0.07}$ & ${2.87}_{-0.15}^{+0.27}$ & 
$2.00$ & $-6.26\mathbf{\pm0.03}$ & $ 4.2\mathbf{^{+1.1}_{-1.0}} \,(\mathbf{1.6^{+0.3}_{-0.4}})$ & $0.35$ \\

\hline
\multicolumn{12}{c}{\bf Type-2 AGN (NGC 1068 SED)} \\
\hline
 & C &
$0.00\,$(\scriptsize{fixed}) &$0.00\,$(\scriptsize{fixed}) & $0.78_{-0.15}^{+0.10}$ &
$2.67_{-0.35}^{+0.41}$ &$44.15_{-0.21}^{+0.20}$ & ${1.79}_{-0.57}^{+0.33}$ & $...$ & 
$-4.77\mathbf{\pm0.03}$ & $9.2\mathbf{^{+6.1}_{-3.3}} \,(\mathbf{3.4^{+1.2}_{-1.6}})$ & $0.18$\\
 & D &
$0.90_{-0.14}^{+0.12}$ &$6.00$ $^{**}$ & $0.00$ $^{**}$ &
$2.66_{-0.25}^{+0.29}$ &$44.11_{-0.16}^{+0.17}$ & ${2.04}_{-0.60}^{+0.70}$ & $...$ & 
$-4.69\mathbf{\pm0.03}$ & $5.5\mathbf{^{+2.6}_{-1.5}} \,(\mathbf{2.0^{+1.0}_{-0.8}})$ & $0.33$\\
\hline
\multicolumn{12}{c}{\bf{Type-2 AGN (Circinus SED)}} \\
\hline
 & E &
$0.00\,$(\scriptsize{fixed}) &$0.00\,$(\scriptsize{fixed}) & $0.76_{-0.14}^{+0.13}$ &
$2.73_{-0.33}^{+0.42}$ &$44.15_{-0.20}^{+0.24}$ & 
${2.50}_{-0.50}^{+0.55}$ & $...$ & $-4.75\mathbf{\pm0.03}$ & 
$14.6\mathbf{^{+8.4}_{-5.1}} (\mathbf{5.4^{+3.1}_{-1.9}})$ & $0.16$\\
 & F &
$0.87_{-0.14}^{+0.12}$ &$6.00$ $^{**}$ & $0.00$ $^{**}$ &
$2.68_{-0.24}^{+0.26}$ &$44.10_{-0.12}^{+0.13}$ & 
${2.55}_{-0.54}^{+0.52}$ & $...$ & $-4.66\mathbf{\pm0.03}$ & 
$7.7\mathbf{^{+3.0}_{-2.1}} (\mathbf{2.9^{+1.1}_{-0.8}})$ & $0.49$\\
\hline \hline
\end{tabular}
\end{center}
\begin{list}{}{}
\item   \scriptsize{$^a$ $L_{15}^*$ corresponds to $\nu L_{\nu}$ at 15$\,\mu$m 
        and is given in $\mathrm{erg \,s}^{-1}$.          
        
        $^b$ The normalisation, $\mathrm{log}\,\Phi^{*}$, is given in 
        $\mathrm{Mpc}^{-3}\,\mathrm{mag}^{-1}$. 
        
        $^c$ Contribution $ \nu I_{\nu}$ to the CIRB at 15$\,\mu$m in units of 
        $10^{-11} \mathrm{W m^{-2} sr^{-1}}$. In parenthesis we report the
        \% of the CIRB produced by the fit given a CIRB value of
        $\mathbf{2.7}$ nW m$^{-2}$ sr$^{-1}$.
        
        $^{**}$ The values of these parameters reached the physical limit
	imposed in our minimisation process. These constraints were imposed to
	the models in order to not produce unphysical solutions at high redshift
	due to the low statistics. Therefore, no errors are reported for these
	parameters.}
\end{list}

\end{table*}

In total, 72 (27 ELAIS + 41 RMS + 4 {\it ISO}-Deep) type-1 AGN were
used to derive the LF.  The best solutions found for each parameter ($\pm
1\sigma$ confidence levels) and the 2D-KS probabilities are presented in Table 2
in the rows labeled "A" and "B" for the fixed and variable faint slope models
respectively.

In Figure\,4 we show the observed space density distribution and the
best fit models in two redshift intervals: $z=[0,0.2]$, where the {\it
IRAS} sources dominate, and $z=[0.2,2.2]$, mainly populated by {\it
ISO} sources. The large interval at high redshifts ($z=[0.2,2.2]$) 
was chosen only for representation purposes, to assure a significant
number of observed sources in each luminosity bin. In
order to correct for evolution within the redshift intervals, the
observed space density distribution is plotted following La Franca et
al. (1997).  The expected number of sources given by the model in each
bin ($N^{model}$) is computed and compared to the observed number of
AGN in the bin ($N^{data}$).  The ratio $N^{data}$/$N^{model}$ for
each bin is then multiplied for the value of the LF at the
corresponding central luminosity and redshift value of the bin.  The
plotted error bars correspond to the 1$\sigma$ Poisson distribution and 
were estimated following Gehrels (1986). Space density upper limits are given
where sources are expected by the model but not observed. The 1$\sigma$
dispersion for the LF was computed according to the 1$\sigma$ uncertainties of
its parameters, and represented as a light-grey shaded area in Fig.\,4.
 
\begin{figure}[!h]
\centering
\resizebox{\hsize}{!}{\includegraphics{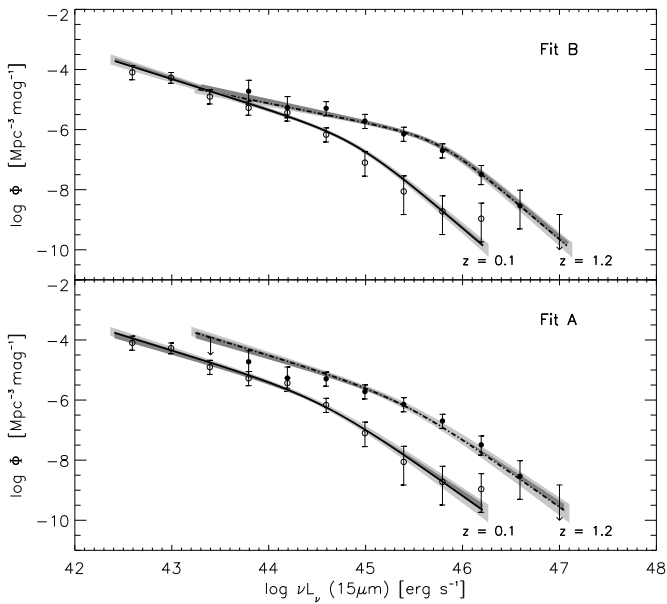}}
\caption{\scriptsize{Type-1 AGN local ($z$=0.1, {\it continuous} line) and 
high redshift ($z$=1.2, {\it dash--dotted} line) best fits to the 15$\,\mu$m 
Luminosity Function assuming a redshift dependent faint slope 
(Fit ``B'', {\it top panel}) and a PLE model (Fit ``A'', {\it bottom panel}). 
Open circles represent the observed space density of the local population 
(dominated mainly by RMS sources) while filled circles represent the 
observed space density of the high redshift population (dominated by ELAIS
sources). Poissonian errors at 1$\sigma$ confidence level are shown. 
The dark-grey shaded areas (quite narrow and thus barely visible) represent
the uncertainty introduced by the assumptions used to estimate the optical 
incompleteness of the samples, while the light-grey shaded areas correspond to
the 1$\sigma$ errors of the best fit parameters of the LF (see \S3.2).}}
\label{fig:type1fit}
\end{figure}

Beside the natural statistical uncertainties due to the limited number of
sources used in our analysis, our fitting technique is affected by possible
errors due the assumed average value of the $L_{15}/L_R$ ratio used to correct
for the spectroscopic incompleteness of the samples. In order to estimate these
uncertainties we have recomputed the LF best fit solutions assuming two
extreme cases for the log($L_{15}/L_R$) ratio. In particular, we have used as
average value of log\,($L_{15}/L_R$) our best estimate (0.23) plus or minus the
observed 1$\sigma$ dispersion (0.25). Even under these extreme assumptions, the
parameters of the LF result to be within the errors quoted in Table 2. The
dark-grey shaded areas in both panels of Fig.\,4 show the corresponding
1$\sigma$ range of uncertainty introduced in the estimate of the space density
of type-1 AGN. As it is clearly seen in the figure, the statistical errors due
to the limited number of sources used in our analysis dominate over the
uncertainties due to the assumptions on the average value of the
log($L_{15}/L_R$) ratio.  For example, the uncertainty introduced in the
estimated density of type-1 AGN by the assumed $L_{15}/L_R$ relation is
$\sim$10\% at $L_{15} \sim L_{15}^\ast$, while the statistical uncertainty due
to the limited number of sources is larger than 20\%.  For these reasons, in
order to compute the errors over the parameters of the LF and any derived
quantity (e.g.  integral counts and redshift distributions), we neglected the
uncertainties related to the assumptions on the average value of the
log($L_{15}/L_R$) ratio.

We observe that although the overall fit by the PLE model (model A in Table 2, 
Fig.\,4 $bottom$ panel) is in reasonable agreement with the observed data, it 
overpredicts by a small amount ($\sim$1$\sigma$), but systematically over 5 
luminosity bins, the number of the low luminosity ($L_{15}(z)<L_{15}^{*}$) 
and high redshift \mbox{($z=1.2$)} type-1 AGN. This is the main reason why a
dependence of the faint slope on the redshift was included.  The introduction
of this dependence translates into a luminosity dependent luminosity evolution
(LDLE) for the faint part of the LF (model B in Table\,2). In this case, a
better agreement between the model and the data is obtained (Fig.\,4, {\it top}
panel) as shown by the increment of the 2D-KS probability (0.35 vs. 0.11).

The measured rate of luminosity evolution ($k_L$) results to be rather
independent from the adopted relation between $\alpha$ and $z$. The best fit
values, $k_L\sim2.7-2.9$, are similar, within the errors, to those already found
for these objects in the optical and X--ray wavelengths, where the evolution
rate is between 2.5 and 3.5 (La Franca et al. 1997; Boyle et al. 2000; Miyaji et
al. 2000 \& 2001; Croom et al. 2004; La Franca et al. 2005; Hasinger et al.
2005). Therefore, according to this analysis, no difference is found in the
evolution of mid-IR selected type-1 AGN in comparison to the ones selected in
the optical and X--rays.

\begin{figure*}[!t]
\centering
\resizebox{\hsize}{!}{\includegraphics{matute05_fig5a.ps}
                      \includegraphics{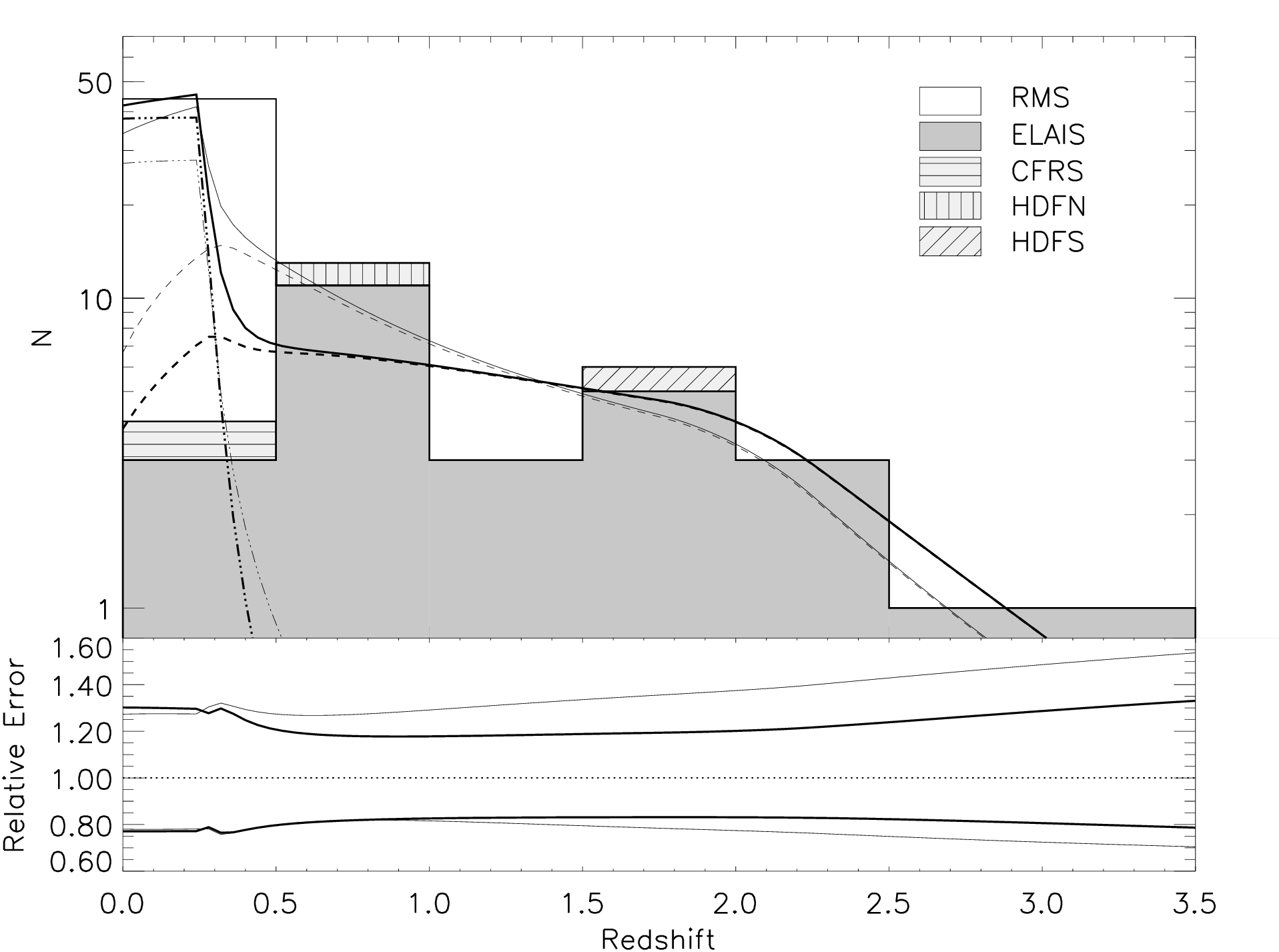}}
\caption{\scriptsize{{\it Left}\,) Mid-IR integral counts for type-1 AGN. 
Shaded areas show the RMS and ELAIS observed counts, while symbols (triangle, 
star and diamonds) represent the {\it ISO}-Deep surveys. The thick {\it solid}
line gives the predicted counts from the variable slope fit model (``B''). 
Expectations from the fixed faint slope model (``A'') are plotted as a {\it
dashed} line. All errors in the observed distributions are quoted at the
1$\sigma$ level and given by Poisson statistics. The bottom panel show the
1$\sigma$ relative error on the predicted counts due to the uncertainties on the
best fit parameters of the LF (see text). Line types have the same meaning as in
the above panel. {\it Right}\,) Redshift distribution for type-1 AGN. The
histogram represents the total number of observed sources in each redshift
interval. Shaded regions outline the ISOCAM contribution, while the open
histogram accounts for the contribution from {\it IRAS} sources. The lines are
the predictions from the best fit model (``B''): thick {\it dash-dotted} for
{\it IRAS}, thick {\it dashed} for {\it ISO} population, while the {\it
continuous} thick line gives their sum.  The expected distributions for the ``A"
model are plotted as above but using thin lines. The bottom panel shows the
relative errors of the predictions for the entire mid-IR population ({\it
IRAS+ISO}) only.}}
\label{fig:zdistrib_counts_agn1}
\end{figure*}

The 15$\,\mu$m integral counts and the observed redshift distribution
for type-1 AGN can now be compared with the predictions derived from
the best fit models.  The results of this comparison are presented in
the upper right and left panels of  Figure\,5. The bottom panels show the
corresponding relative errors as derived from the 1$\sigma$ uncertainties on 
the best fit parameters of the LF.

The expected integral counts (Fig. 5, $left$) given by both models
({\it dashed} line for model ``A'' and {\it continuous} line for model
``B'') provide a good representation of the observed data (shaded
areas) down to $\sim$2 mJy. At fluxes below $\sim$2 mJy the variable
slope model agrees better with the ELAIS data, while the PLE model is
more representative of the fainter population in the HDF and CFRS
fields. At the faintest fluxes ($S_{15}=0.1$ mJy) the variable slope
model underpredicts the observed counts by a factor $\sim$2. 
The uncertainties on the predicted counts at this flux ($\sim$20\%)
are not large enough to justify the observed discrepancy.

We also find an overall good agreement between the redshift distribution 
of the observed data and the model predictions (Fig.5, $right$).
The largest discrepancy between the models occurs at the lowest
redshift bin ($z=[0,0.5]$) where the fixed faint slope model (thin
lines) overpredicts the number of sources observed in the ISOCAM
fields by a factor $\sim$3 (4 observed vs. $\sim$12 expected), while
the variable faint slope model provides a better representation of
the observed distribution.

In summary, although the variable faint slope model provides a better
overall fit to the LF and reproduces better the $z$-distribution, it
underestimates the faint mid-IR counts \mbox{($S_{15}<0.5$\,mJy)} by a factor
of $\sim$4. As both models (``A'' fixed \& ``B'' variable faint slope)
are statistically acceptable, we conclude that given the presently 
available statistics the two models are equivalent within the errors.

\begin{figure*}[!ht]
\centering
\resizebox{\hsize}{!}{\includegraphics{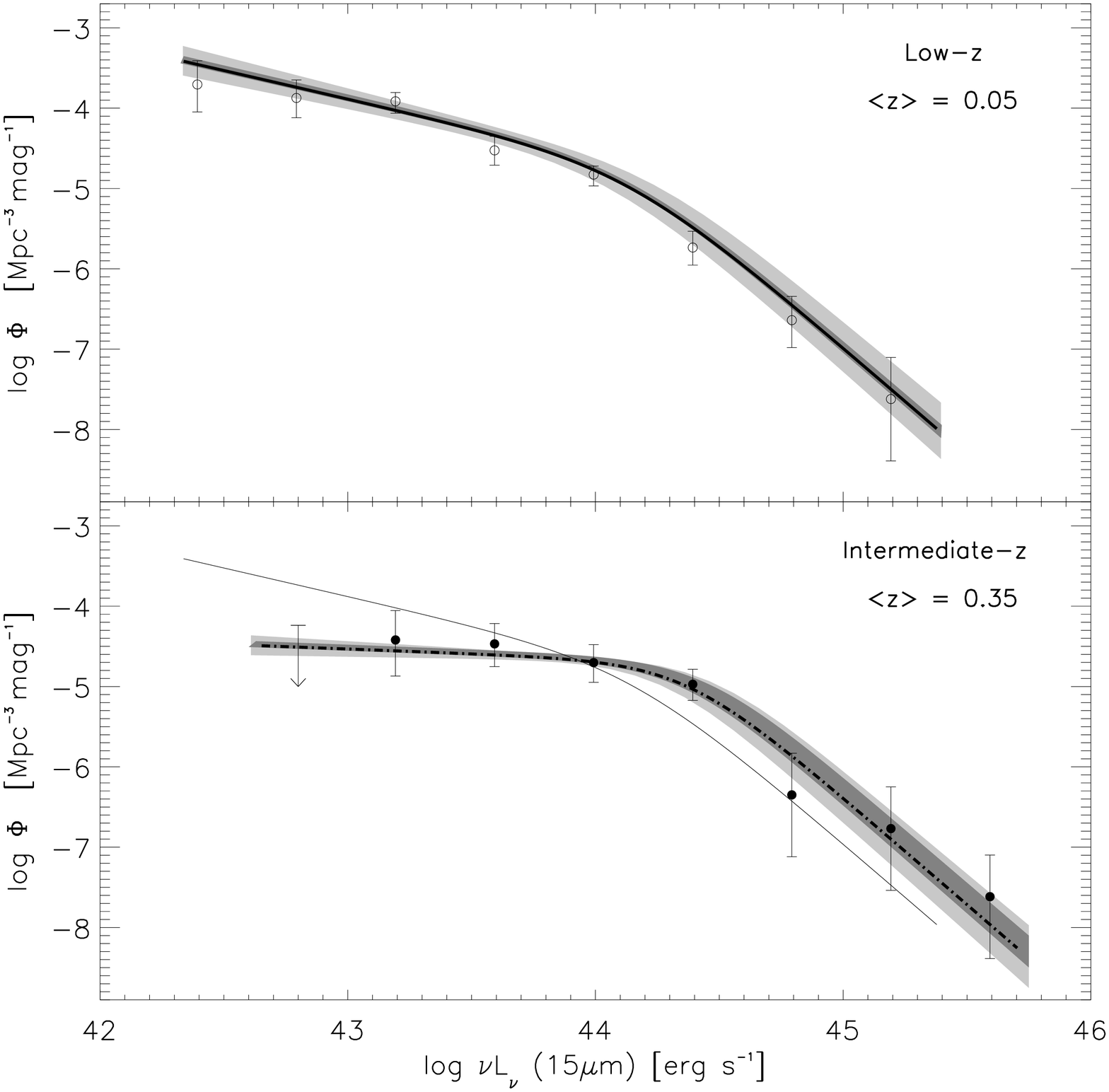}
                      \includegraphics{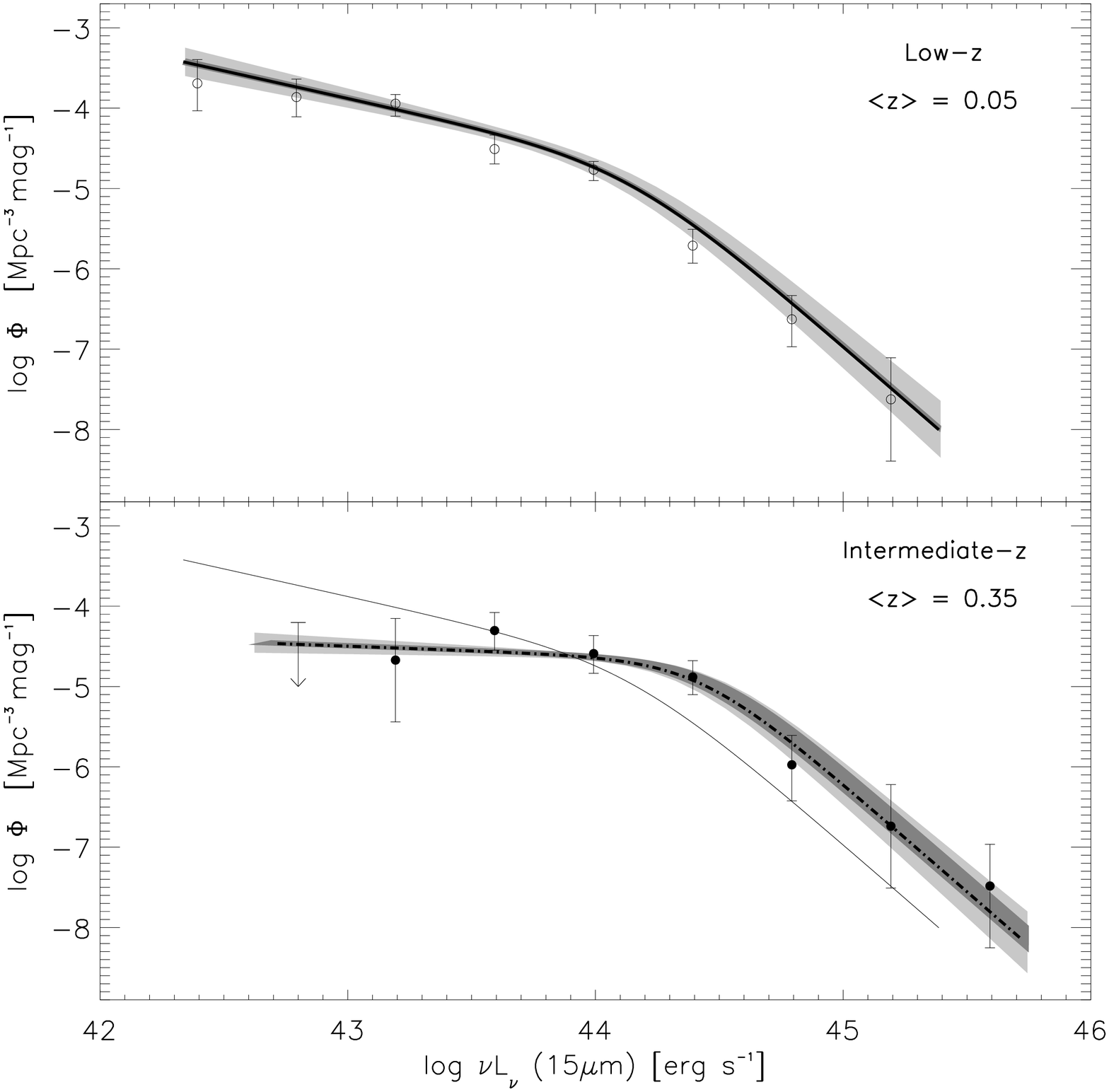}}
\caption{\scriptsize{Observed and best fit LF for type-2 AGN  at
    $<z>$=0.05 ($continuous$ line) and  $<z>$=0.35 ($dash$--$dotted$
    line). The solutions with a redshift dependent faint slope, assuming a
    NGC\,1068 (``D''; $left$ panel) and Circinus (``F''; $right$ panel) 
    $k$-corrections, are shown. Symbols and shaded areas as in Fig. {\bf 4}. 
    Due to the small redshift intervals used, and for the sake of clarity, the
    plots have been divided into two panels for the low-redshift ({\it
      top}) and intermediate-$z$ sources ({\it bottom}).  At
    intermediate-$z$, the thin line shows the low-$z$ best fit.}}
\label{fig:type2fit}
\end{figure*}

M02 measured for the first time the evolution of type-1 AGN based on a
preliminary mid-IR catalogue from the ELAIS-S1 field. The values found
here for the fixed faint slope LF (model ``A'') are in agreement
with what already found by M02 within $\sim$1$\sigma$ errors. The
only exception is for the bright slope ($\beta$), where the here derived value
is flatter (2.29$^{+0.36}_{-0.16}$ vs. 2.89$^{+0.29}_{-0.26}$).  
This difference can be understood if we take into account the large 
uncertainties in the brighter luminosity intervals (in both M02 and our sample) 
due to the small number of sources.  While the 2D-KS test gives a 11\% 
probability that the our sample of AGN is well represented by a non-evolving 
faint slope (model ``A''), this probability decreases to 5\% if we use the
parameters values reported by M02 (Table\,1 in M02).

\subsection{The Evolution of type-2 AGN}

As for type-1 AGN, a relation between the intrinsic 15$\,\mu$m luminosity
($L_{15}$) and the optical luminosity ($L_{R}$) for type-2 AGN is required to
correct for the optical incompleteness. FLF04 has shown that such a relation
exists and found it to be valid for the whole mid-IR population (excluding
type-1 sources) over more than 4 decades of mid-IR luminosity. In order to
derive this relationship, optical identifications from RMS, ELAIS-S fields, HDF
North (Aussel et al. 1999a, 1999b) and South (Mann et al. 2002; Franceschini et
al. 2003) were used.  As opposed to type-1 sources, the ratio $(L_{15}/L_R)$ has
a dependence on $L_{15}$ which can be expressed in the linear form
$${\mathrm {log}}(L_{15}/L_R)= 0.47 \, {\mathrm {log}}L_{15} - 5.02,$$
where luminosities are expressed in solar units (see FLF04, Fig.\,13).  The data
show an intrinsic 1$\sigma$ dispersion of 0.32. This equation implies that
$L_{15}$ increases $\sim$3 times faster than $L_R$. The relation works for both
nearby and distant objects (from $z$=0 up to $z$=0.7) and, therefore, has been
used to estimate the fraction of sources lost due to the spectroscopic limit of
the different surveys, i.e. the $\Theta(z,L_{15})$ term used in the minimisation
of the S function (see eq.\,3).

The LF was computed in the luminosity range log$L_{15}=[42,47]$ and in the 
redshift interval $z=[0,0.7]$. A total of 75 sources (25 from ELAIS and 50 
from RMS) were used to derive the LF shape and evolution. Best fit values for 
the LF parameters, as well as the corresponding 2D-KS probabilities, can be 
found in Table\,2 for each of the two representative mid-IR $k$-corrections 
assumed.

For type-2 AGN, we followed the same fitting sequence as for type-1
AGN.  The results for a fixed faint slope (PLE model) are found on
Table 2 on rows labeled ``C'' (for the NGC 1068 $k$-correction) and
``E'' (for the Circinus $k$-correction).  In both cases, the value of
the 2D-KS test (P$_{2DKS}$=0.16 \& 0.18 respectively) does not reject
the PLE model, but an important source overprediction is again present
in the model if compared to the observed data for the low luminosity
and moderate-$z$ ($z$=[0.1,0.6]) part of the sample. A better fit is
found by assuming an evolving faint slope of the LF parameterised as
specified in $\S$3.2.  In this case, in order to avoid unphysical
solutions due to the low statistics at high redshift, the $\alpha_2$
parameter was allowed to vary in the range $[0,6]$, while the $\alpha_3$
parameter was allowed to vary in the range $[0,2]$. Indeed, both parameters
found a solution at the edges of the allowed ranges. 
The results can be found in Table 2 and Figure 6, fits labeled ``D'' 
and ``F'' for the NGC 1068 and Circinus $k$-corrections respectively.

\begin{figure*}[!t]
\centering
\resizebox{\hsize}{!}{\includegraphics{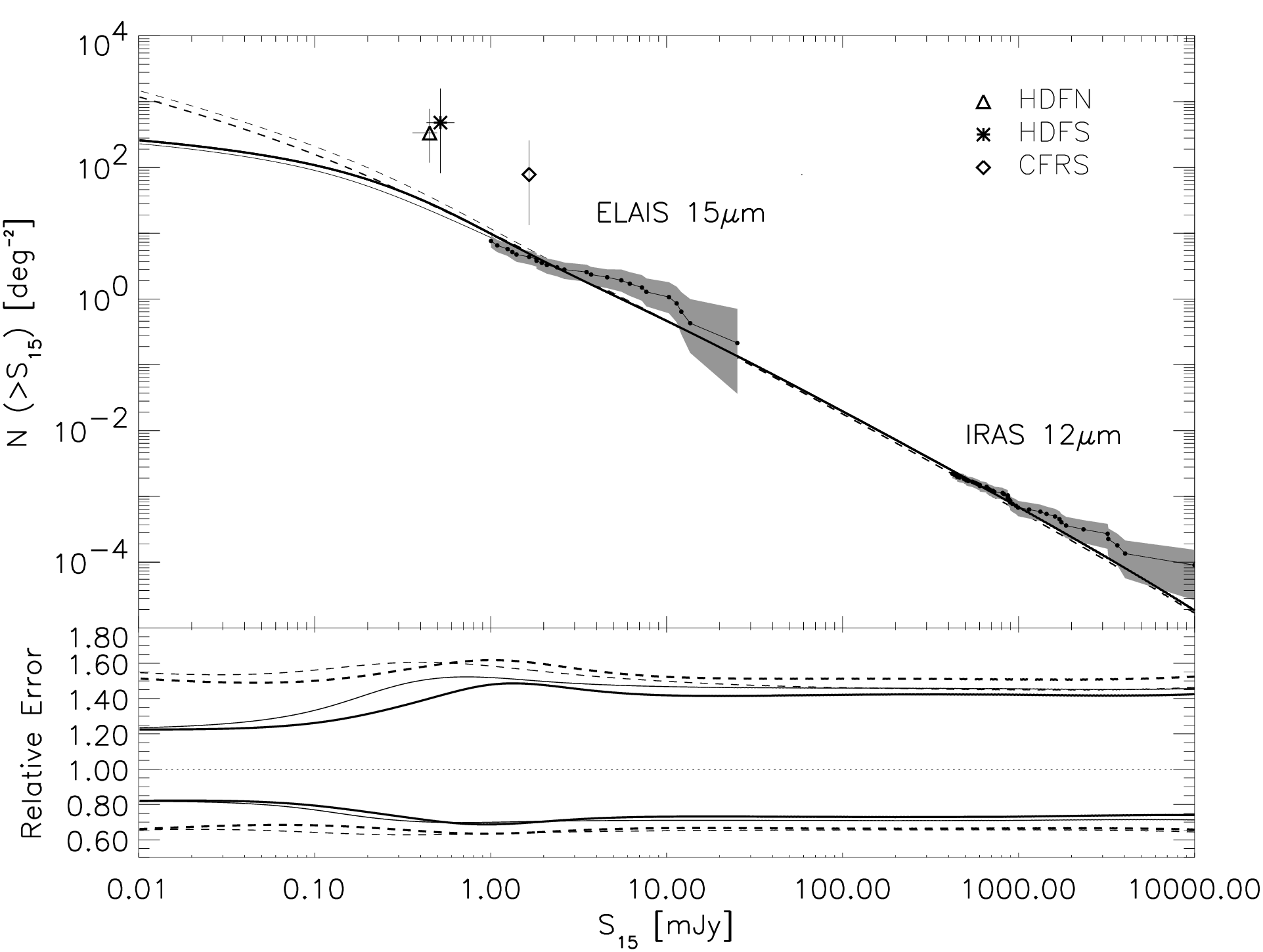}
                      \includegraphics{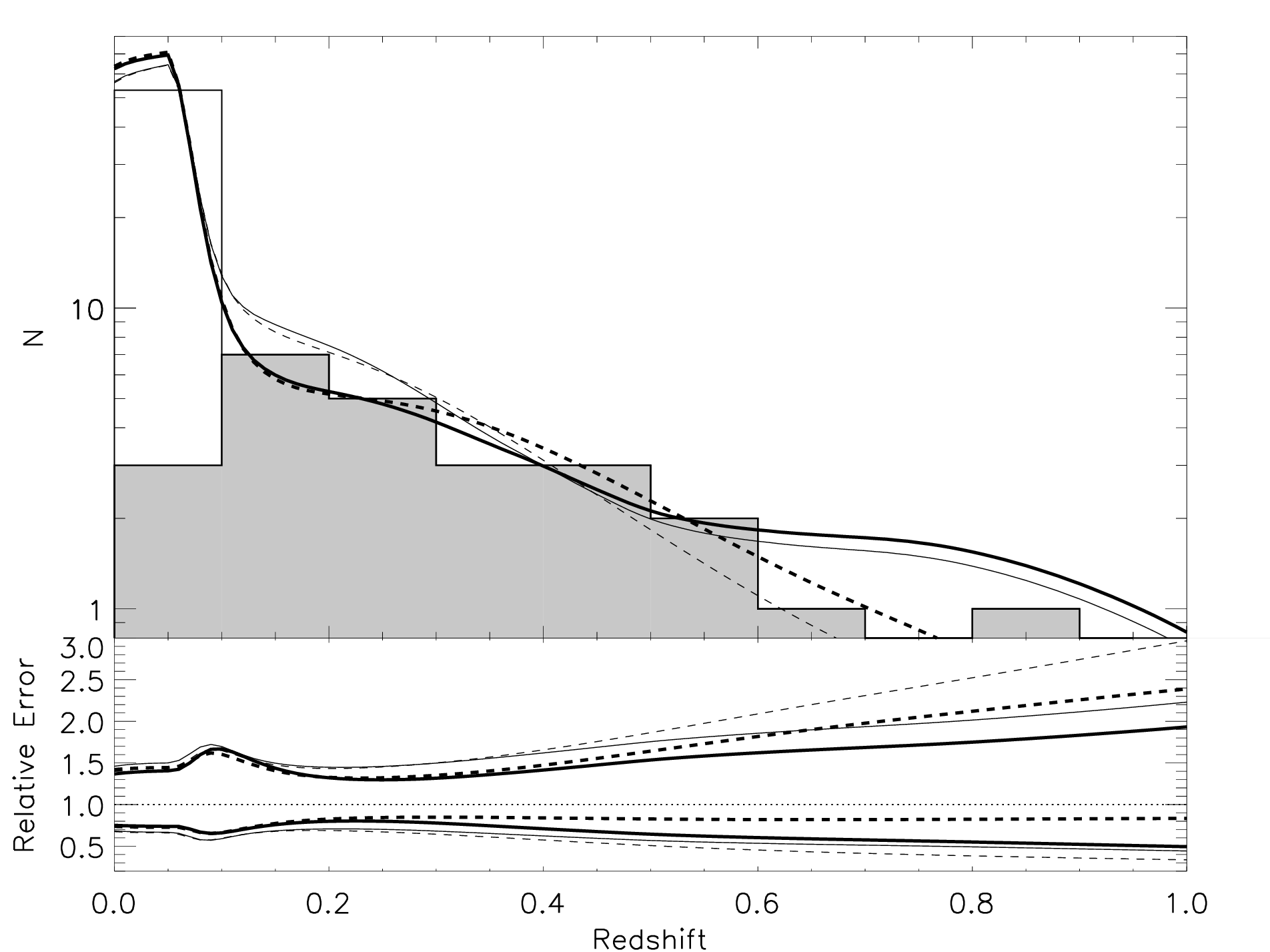}}
\caption{\scriptsize{{\it Left\,}) Observed and predicted integral counts for 
type-2 sources using the the two representative SED: NGC\,1068 (thin-$dashed$ 
for the ``C'' model and thin-$continuous$ for the ``D'' model) and Circinus (thick-$dashed$ for the ``E'' model and thick-$continuous$ for the ``F'' model). 
{\it Right}\,) Observed and predicted type-2 AGN redshift distribution for the 
different models mentioned before. Lines as in $left$ panel. Symbols and 
shaded area in both panels as in Fig. 5. Errors reported in the lower 
section of both panels have been computed and plotted as described in Fig.\,5.}}
\label{fig:zdistrib_counts_agn2}
\end{figure*}

Fig.\,6 shows the observed and predicted space density distribution of
the type-2 AGN in two representative redshift intervals, {\it low-$z$}
($z=[0,0.1]$) where the {\it IRAS} sources dominate and {\it
  intermediate-$z$} ($z=[0.1,0.6]$),  mainly populated by {\it ISO} 
sources. The overall best solution found for the two representative SED,
NGC\,1068 ($left$ panel, model ``D'') and Circinus ($right$ panel,
model ``F''), are plotted. The observed space density, errors, 
upper limits and the 1$\sigma$ confidence regions for the LF have
been computed and plotted as described in \S3.2.  As observed for
type-1 AGN, the errors introduced by the low statistics (light
shaded area) are larger than the ones due to the assumptions on the
$L_{15}/L_R$ relation (dark shaded area). For example, in this case,
while the error introduced by the assumed $L_{15}/L_R$ relation in
the determination of the density of type-2 AGN is $\sim$20\% at the
break luminosity ($L_{15}^*$) of the LF, the limited statistics
introduce, instead, a $\sim$60\% uncertainty. Therefore, also for
type-2 AGN, in order to compute the errors on the LF parameters and
on any derived quantity (counts and redshift distributions) the
uncertainties introduced by the assumptions on the value of the
$L_{15}/L_R$ relation have been neglected.

The rate of evolution $k_L$ found for type-2 AGN depends strongly on
the assumed SED.  If an NGC\,1068 $k$-correction is adopted, the rate
of evolution is $k_L\sim2.0\pm0.5$, lower than the one measured for
type-1 sources ($k_L\sim2.8\pm0.3$).  On the other hand, an evolution
rate $k_L\sim2.5\pm0.5$ is obtained if the Circinus SED is assumed,
closer to the one measured for type-1 AGN. These  differences in $k_L$
are caused by the stronger $k$-correction introduced by the Circinus SED 
(Fig. 1), which generates higher intrinsic luminosities in the redshift range
[0.1,0.6] compared to those given by the NGC\,1068 $k$-correction. At variance,
the impact of the assumed SED on all the other fitted parameters is small (see
Table 2 \& Fig. 6).

For each best fit LF of the type-2 AGN, the predicted counts and
redshift distributions were derived and compared with the observed
data. Predictions at 15$\,\mu$m for the integral counts are given in
the $left$ panel of Fig. 7, while the derived redshift distribution is
shown in the $right$ panel of the same figure.  In both panels of
Fig.\, 7 a thick-$dashed$ line represents the Circinus fixed slope
model (``E''), a thick-$continuous$ line the Circinus variable slope
model (``F''), a thin-$dashed$ line the NGC\,1068 fixed slope model
(``C'') and a thick-$continuous$ line the NGC\,1068 variable slope
model (``D'').

A good agreement, within the errors, between the models and the data
is observed.  At variance with the solutions found for the type-1 AGN
(\S 3.2), although a higher 2D-KS probability is obtained using a
variable faint slope model, no major differences are observed in the
integral counts down the faintest ELAIS flux ($\sim$1\,mJy;
log$S_{15}\sim0$).  The differences between the various models only
become significant at fainter fluxes ($S_{15}\le 1$\,mJy), especially
at the fluxes reached by the {\it Spitzer} space observatory
($S_{24}\sim 0.01$\,mJy; see \S 6).  Indeed, the relative errors
are rather uniform, with mean values around 30-50\%. Therefore, the
higher observed density of the deep {\it ISO} data (a factor $\sim$10) 
is statistical significant. However, as previously mentioned
(\S2), the higher density of AGN found in these fields is partly due to
the classification method used, which is not only based on a pure
optical spectral classification. The predicted redshift distributions
agree, within the errors, in the redshift interval where the LF was
computed ($z=[0,0.7]$).  The different behaviour observed at $z>0.7$
is understood by the different $k$-corrections obtained from the
adopted SED.  The estimated errors range from a factor $\sim$1.3
at low redshift to $\sim$2.0 at the limiting redshift used to fit
the LF, and increase drastically when the predictions are extrapolated at
higher redshifts.

\subsection{The missing fraction of AGN}

In the previous sections we have seen that a flattening with redshift of the
faint slope of both type-1 and type-2 luminosity functions is  slightly
favoured by the data. This flattening would disappear if $\sim$7 additional
type-1 AGN, in the luminosity-redshift range log$L_{15}$=[42,45] and
$z$=[0.2,3.2], and $\sim$6 additional type-2 AGN within log$L_{15}$=[42,44] and
$z$=[0.1,0.7] were found. We briefly discuss now two possible origins of
incompleteness for the mid-IR AGN sample and their effects on the observed LF:
i) objects not identified due to the spectro-photometric limit of completeness
of the survey and, ii) possible misclassification of the objects.
 
The spectroscopic incompleteness at faint optical magnitudes of the
ELAIS sample is represented by the factor $\Theta(z,L_{15})$ (\S 3.1).
The corrections introduced through this term are represented in Fig. 8
as a function of redshift and mid-IR luminosity.
The shaded areas in the figure outline the region where type-1 ($top$)
and type-2 sources ($bottom$), fainter than the spectro-photometric
limit of the surveys, are expected to be found. A darker area in the
plot indicates that a higher fraction of correction has been applied 
(i.e. higher number of lost sources) and it spans from $\sim$1\% 
(light-grey area) to $\sim$30\% (darkest area) of the total sky area 
available in a given interval of redshift and luminosity.  The correction 
computed corresponds to $\sim$4 type-1 and $\sim$1 type-2 sources in 
the $L_{15}-z$ interval used in the minimisation process. 
If the suggested flattening is really due to an incompleteness effect, then the
possibly missed objects are expected to follow an optical-mid-IR relation
different from the one measured for the identified fraction of the
high redshift {\it ISO} sources and in the local universe by {\it IRAS}
($z<0.1$) and adopted in our estimate of $\Theta(z,L_{15})$. 
In the case of type-1 AGN, if the average value of the assumed 
log($L_{15}/L_R$) relation is increased by 2.5 times the observed dispersion
of the relation (i.e. $\sim$0.86 instead of 0.23), $\sim$7 additional sources 
would be expected at low luminosity and high redshifts, thus avoiding the 
need for a flattening of the LF. This higher $L_{15}/L_R$ value could correspond, 
for example, to sources with a significantly larger amount of gas and dust 
with respect to the local samples of Elvis et al. (1994) and Spinoglio et al.
(1995). For type-2 AGN, the relation derived by La Franca et al. (2004)
implies that for the low luminosity ($L_{15}<L^*_{15}$) and
moderate redshift ($z\sim$0.1--0.5) sources, all the possible optical
counterparts should have been observed. Only the high luminosity
($L_{15}>L^*_{15}$), higher redshift ($z>0.5$) sources are affected by
incompleteness. Also in this case a significant higher value of the 
$L_{15}/L_R$ relation would be required in order to avoid the need 
for a flattening of the LF.

\begin{figure}[!ht]
\centering
\resizebox{\hsize}{!}{\includegraphics{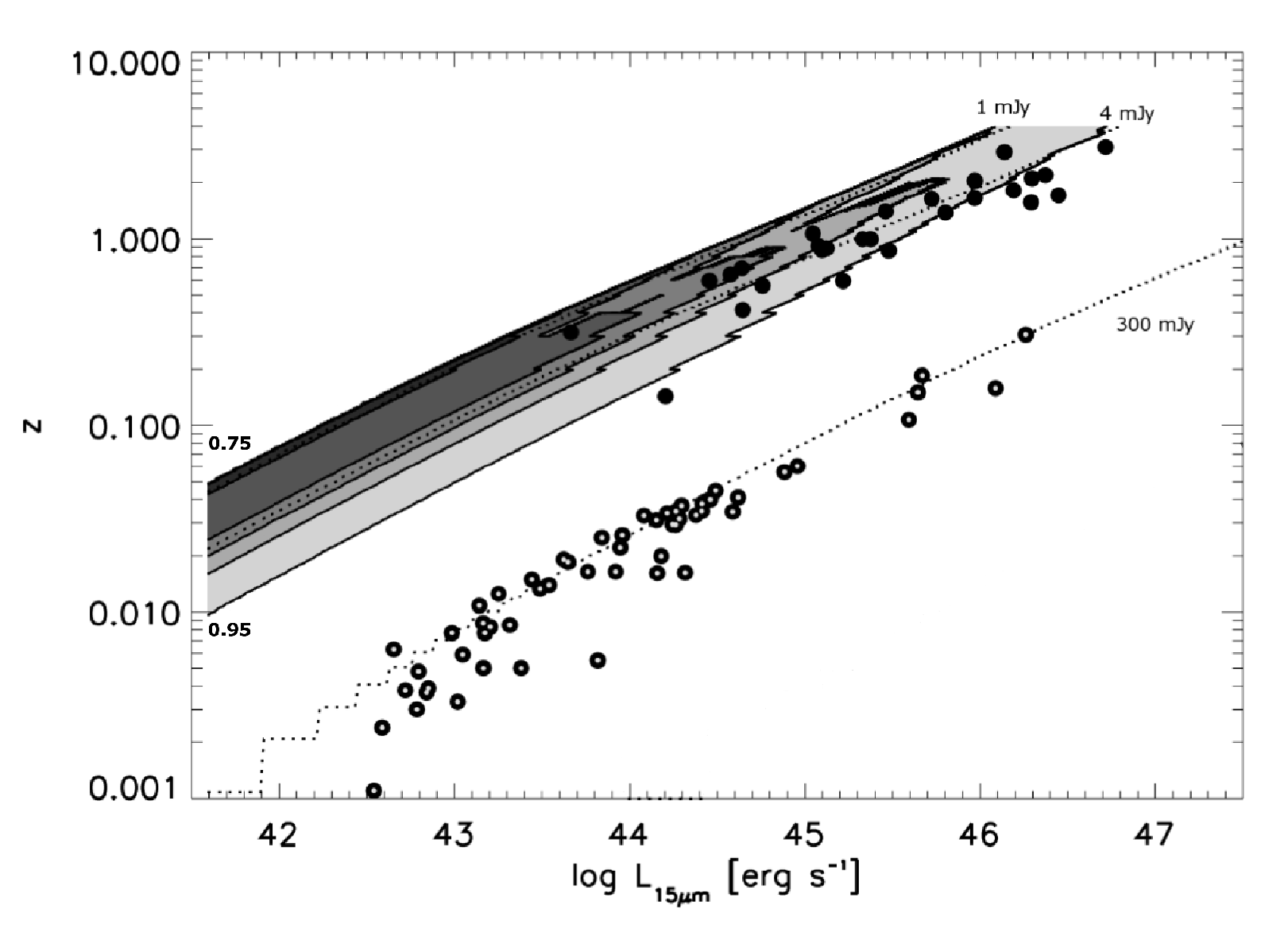}}
\resizebox{\hsize}{!}{\includegraphics{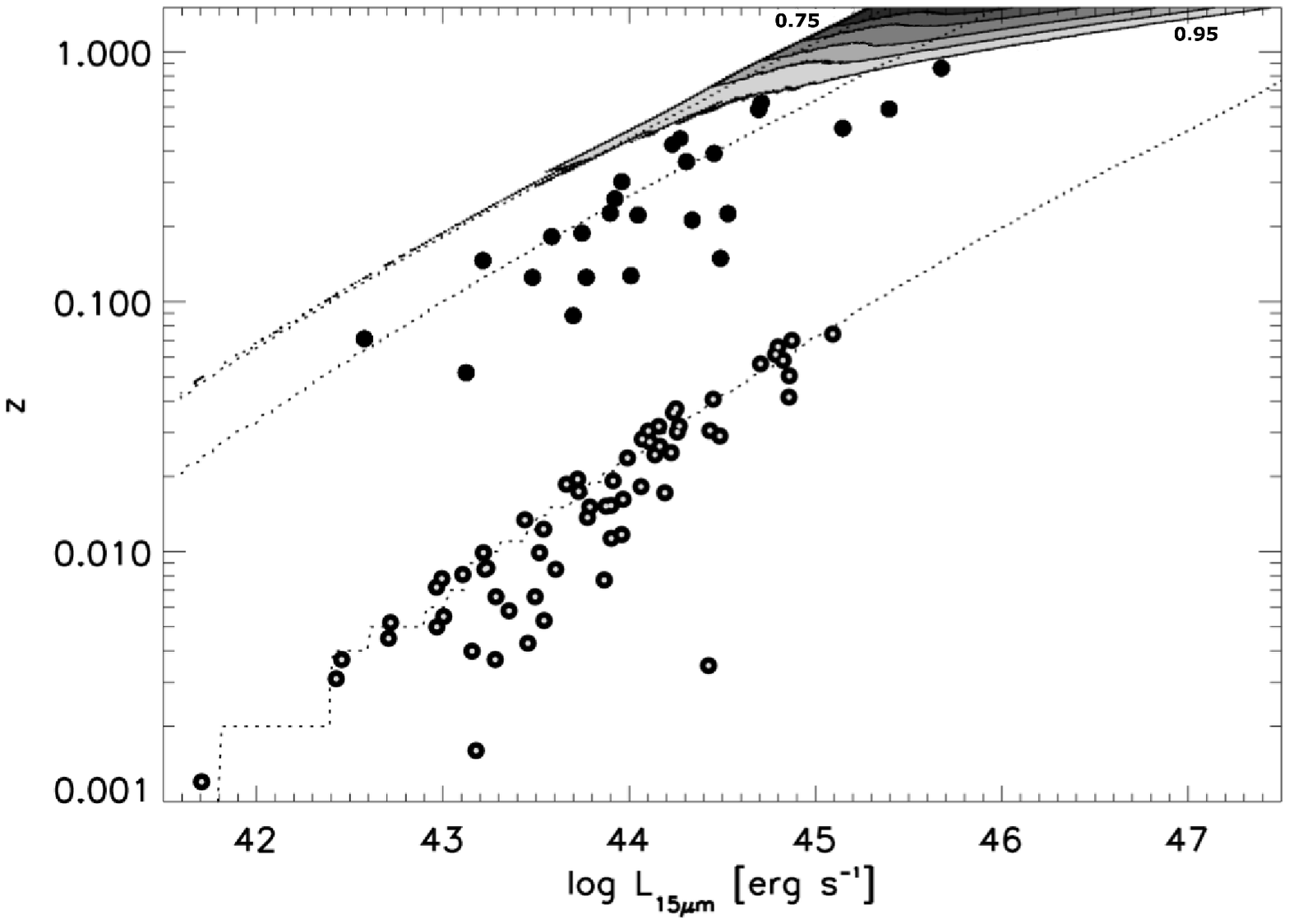}}
\caption{\scriptsize{Luminosity--redshift distribution of the type-1 
({\it top}) and type-2 ({\it bottom}) AGN. Shaded areas represents 
the fraction of incompleteness due to the optical spectroscopic limits. 
 Contours range from 75\% (black) to 95\% (light-gray) of the total
area lost and have a 5\% step between them. For type-1 AGN $\sim$4 missed
sources are expected within $z$=[0.1,1.2]. Most missed type-2 AGN are only
expected with $\mathrm {log}\,L_{15}>44.5$ at $z>0.7$, a redshift range outside 
the current analysis (see text). Only $\sim$1 missed type-2 source is expected
to lie within $z=[0.3,0.5]$ and $L_{15}=[43.5,45.0]$. Dotted lines in both
panels show three representative flux limits at 1, 4 and 300 mJy.}}
\label{fig:opt_correction1}
\end{figure}

A second source of incompleteness could be caused by a spectroscopic
mis-classification of the sources. The increase in the mean redshift
of the {\it ISO} population with respect to the local {\it IRAS}
sample (see Table\,1) implies that a larger fraction of the host
galaxy light is collected for a given size of the slit used for the
optical spectroscopic identifications. The consequence is a dilution
of AGN signatures that affects the measured equivalent widths (EW) and
line ratios (see e.g. Moran et al.  2002). The effect is less
important for type-1 sources due to the presence of easily identified
broad emission lines caused by a direct view of the central energy
source with a low level of obscuration.  The situation is more
complicated for type-2 AGN since their higher degree of obscuration
suppresses an important fraction of the optical light from the nuclei.
The host galaxy would then naturally contribute with a larger fraction
to the observed optical light from the nuclear source, sometimes even
dominating it. The effect is maximised for the intrinsically faint
AGN, whose optical features will be highly diluted into the host
galaxy spectrum. This effect could be relevant in producing the
observed flattening at high redshift of the luminosity function.

Observations with {\it XMM} and {\it Chandra} have uncovered a not
negligible number of AGN for which the classification given by their
optical spectra does not correspond to the X--ray classification of
the source (e.g. Fiore et al. 2000, 2003; Barger et al. 2005).  In
their analysis of deep ($\sim$70 Ksec long) {\it XMM} observations on
ELAIS S1-5, La Franca et al. (in preparation) find that the fraction
of misclassified type-1 AGN appears to be negligible. Vice-versa, in
the case of type-2 AGN, the observations with {\it XMM} have revealed
that about 10-20\% of the {\it ISO} counterparts optically classified
as no-AGN (i.e. normal and starburst galaxies) show X--ray luminosity
consistent with AGN activity. This fraction, in the case of ELAIS-S,
would imply that as many as 10 to 25 sources, classified in the
optical as starburst or normal galaxies, may harbour an AGN that can
contribute to the observed mid-IR flux. If this is the case, the
number of true type-2 AGN may be higher by a factor $\sim \!1.4-2.0$
than that used in this analysis and this would have important
consequences for the observed LF and its evolution.

We conclude that the observed flattening with redshift of the LF for
low luminosity AGN can be explained if we assume that the
incompleteness in the ELAIS sample was not properly modeled, for two
possible reasons: (i) the unidentified fraction of sources may have a
$L_{15}/L_R$ ratio different from that of the identified AGN or, (ii)
more likely, a relevant fraction of type-2 AGN could have been
mis--classified due to dilution of the optical nuclear spectra by the
hosting galaxy.  These sources of incompleteness are probably
almost negligible for type-1 AGN (especially the last one), while
our measure of the density of type-2 AGN has instead to be
considered a lower limit.

\section{Comparison with previous works}

\subsection{The {\it IRAS} 12$\,\mu$m local luminosity function}

A local luminosity function (LLF) for type-1 and type-2 AGN was derived at
12$\,\mu$m by RMS. Their results can be directly compared to our findings at
$z$=0. The values for the LF parameters reported by RMS and those computed by us
are summarised in Table 3. 

Two parameters of the LF computed by RMS, the faint slope ($\alpha$) and the
luminosity break ($L_{15}^*$), differ significantly from our solutions for both
the type-1 and the type-2 AGN.  The difference on the value of the luminosity
break can not be explained by the SEDs conversion factors from 12 to 15$\,\mu$m,
since they imply a small change in the intrinsic luminosity of the sources
($\Delta$ log$L[12-15\,\mu$m] $\sim$0.0 for type-1 and $\sim$0.1 for type-2
AGN). These discrepancies can instead be understood if we take into account
that: (i) the faint slope was fixed {\it a priori} in the fitting procedure by
RMS; (ii) the strong effect of evolution in the different luminosity bins was
not considered by RMS and, (iii) our best fit is obtained by taking into account
not only the local {\it IRAS} sources but also the high redshift ISOCAM
population.

\begin{table}
\caption[]{Parameter values of the fit of the mid-IR Luminosity Functions}
\label{Table3}
        \begin{tabular}{lccccccccc}
        \hline
 & \multicolumn{3}{c}{\bf \scriptsize{Type-1}} &\multicolumn{3}{c}{\bf \scriptsize{Type-2}} & 
        \multicolumn{3}{c}{\bf \scriptsize{All}} \\
 & {\scriptsize {$\mathrm{\alpha}$}} & {\scriptsize  $\!\beta$}  
        & {\scriptsize  $\! \mathrm{log}L_{15}^{*}$} & {\scriptsize $\alpha$} & {\scriptsize $\beta$} 
        & {\scriptsize $\mathrm{log}L_{15}^*$} & {\scriptsize $\alpha$} &  {\scriptsize $\beta$} & 
        {\scriptsize $\mathrm{log}L_{15}^*$} \\
\hline
\scriptsize{RMS}$\,^{(a)}$ & \scriptsize{0.0} & \scriptsize{2.1} & \scriptsize{42.8} & \scriptsize{0.0} & 
        \scriptsize{2.5} & \scriptsize{43.2} & \scriptsize{0.3} & \scriptsize{2.1} & \scriptsize{43.4} \\
\scriptsize{Xu03}$\,^{(a,b)}$ & - & - & - & - & - & - & {\scriptsize 0.3} & {\scriptsize 1.7} & {\scriptsize 43.4} \\
\scriptsize{Here}\scriptsize{(low-z)} $^{(c)}$ & {\scriptsize 1.3} & {\scriptsize 2.8} & 
        {\scriptsize 44.8} & {\scriptsize 0.9} & {\scriptsize 2.7} & {\scriptsize 44.1} & - & - & - \\
\scriptsize{Here}\scriptsize{(int-z)} $^{(c,d)}$ & {\scriptsize 0.7} & {\scriptsize 2.8} & 
        {\scriptsize 45.4} & {\scriptsize 0.0} & {\scriptsize 2.7} & {\scriptsize 44.6} & - & - & - \\
\hline
\end{tabular}
\begin{itemize}{}
\tiny{
\item[] $^{(a)}$ Luminosities have been converted to 15$\,\mu$m using the 
		adopted SEDs.
\item[] $^{(b)}$ From Xu et al. 2003.
\item[] $^{(c)}$ Best fit solutions for type-1 (model ``B'') and type-2 
		(Circinus SED, model ``F'').
\item[] $^{(d)}$ Intermediate redshift, $z$=0.6.
}
\end{itemize} 
\label{tbl:MidIR_LF_literature}
\end{table}

\subsection{The {\it IRAS} 25$\mu$m combined AGN sample}

A local combined (i.e. type-1 + type-2) AGN LF has been derived by Xu
et al.  (2003) at 25$\,\mu$m. It is based on an {\it IRAS} colour
selected ($S_{60\mu m}/S_{25\mu m}<5$) sample obtained from a complete
catalog compiled by Shupe et al. (1998). Assuming a luminosity
evolution equal to the optical one (Boyle et al. 2000) and a set of
SEDs with a dependence on luminosity, Xu et al. (2003) were able to
derive the predictions for the AGN counts, the redshift distribution and
the contribution to the cosmic background from the UV to the sub-mm
wavelengths. A direct comparison of the two LFs can then be made by
translating our 15$\,\mu$m luminosities into 25$\,\mu$m luminosities,
using the adopted SEDs for each source class.
The results of this comparison, in two different redshift intervals,
can be found in Fig.\,9 and Table\,3.  We find a good agreement between the
two LFs ({\it dashed} line for the 25$\,\mu$m selected sample; {\it
continuous} line for our type-1 + type-2 15$\,\mu$m LF) in the local
universe ({\it z}$\sim$0.1).  The difference found at intermediate-$z$
($z$=0.6, the maximum $z$ used to derive the type-2 LF) and low
luminosities $L<L^*$) is mainly due to the observed decline in space
density at this redshift for the obscured sources. 
The difference at higher luminosities ($L>L^*$) can be explained by
the stronger evolutionary rate found by Boyle et al. (2000) for
optical selected type-1 AGN ($k_L \sim 3.5$) and used by Xu et al. (2003) 
in their analysis with respect to the values found in this work, $k\sim2.0-2.9$
depending on the AGN type. The difference in the LF induced by these
different $k_L$ parameters is already significant at $z \sim 0.6$.

\subsection{ISOCAM Deep observations}

The deep observations carried out by the ISOCAM instrument on-board of
{\it ISO} place interesting constraints on the shape and behaviour of
the LF. For type-1 AGN they favour a PLE model scenario (dashed line in
Fig. 5, $left$ panel). This suggests that a small fraction of the
low-luminosity ELAIS-S type-1 AGN may still be hidden within the rest
of the mid-IR population. If this is the case, they do not follow the
optical-mid-IR luminosity relation assumed for the rest of the type-1
sources as already discussed in \S3.4.

The observations of type-2 sources in these fields indicate that all
our best fit solutions underestimate them by more than 1.5$\sigma$.
This result is not surprising since most of the type-2 sources
identified in these surveys have been classified using criteria
different from the pure optical spectral classification (i.e. SED
reconstruction, X--ray properties and radio emission). Indeed, this was
the reason for which we did not consider these sources in our minimisation
process. A comparison between their observed space density and our
predictions has to be made with care due to the different classification 
methods used.

We would also like to note here that even if cosmic variance can
increase the HDF-N, -S and CFRS quoted errors in Fig. 5 and Fig. 7 (by
as much as 20--100\% in such a small fields, Somerville et al. 2004),
it can not fully justify the observed difference since in all these fields
the observed counts are found to be systematically higher than our
predictions.

\section{Contribution to the CIRB}

\begin{figure}[!t]
\resizebox{\hsize}{!}{\includegraphics{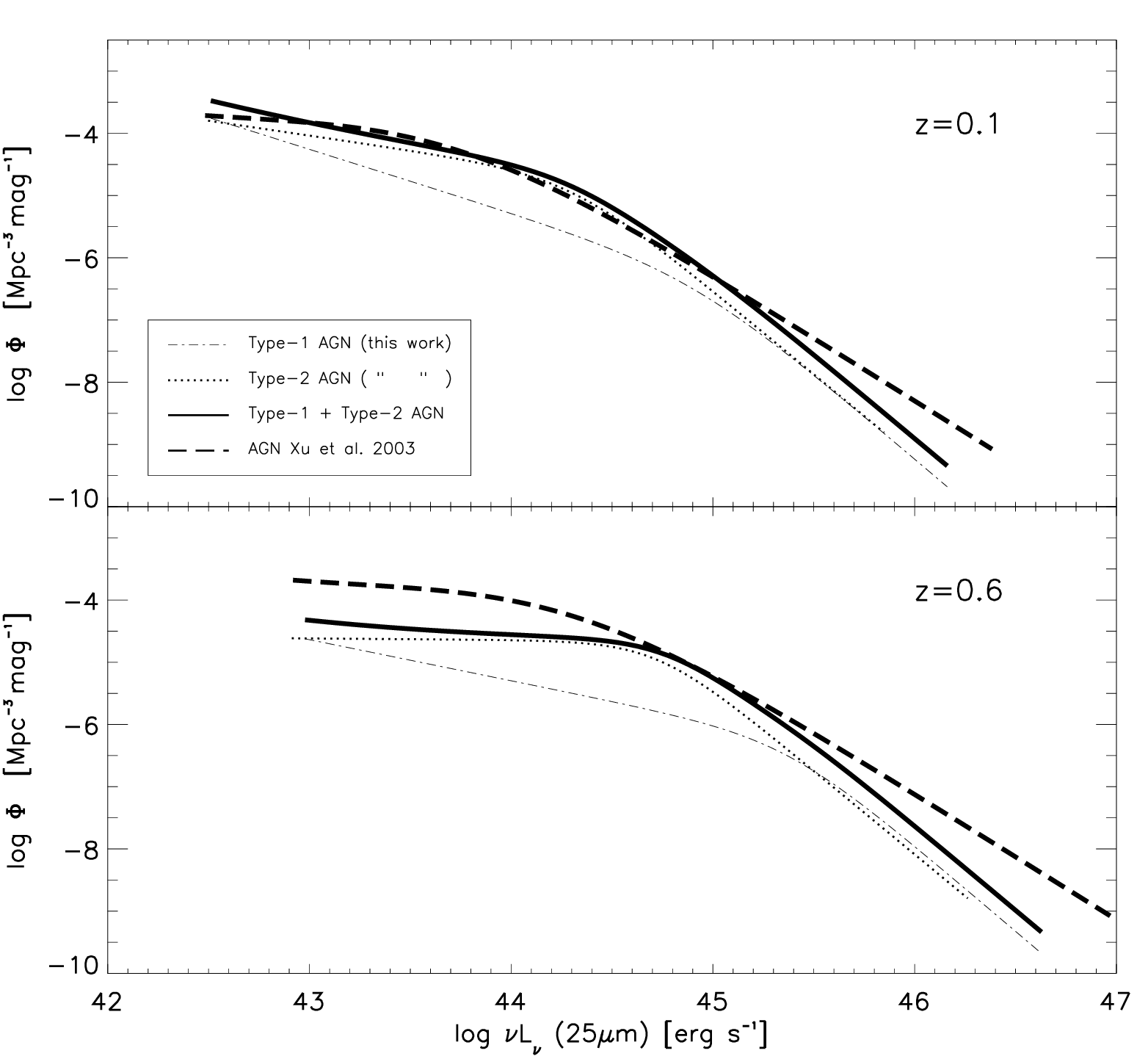}}
\caption{\scriptsize{Comparison of LFs at 25$\,\mu$m. The derived 15$\,\mu$m LF
for type-1 (model ``B'') and type-2 (Circinus SED, model ``F'') AGN, converted
to 25$\,\mu$m luminosities, are compared with the global AGN LF from Xu et
al.(2003). The comparison is made in two different redshift intervals: $z$=0.1
({\it top} panel) and $z$=0.6 ({\it bottom} panel). The values for the
parameters of the LFs are reported in Table 3.}}
\label{fig:combined-sample}
\end{figure}

The contribution of AGN to the intensity of the Cosmic Background
light in the Infrared (CIRB) has been derived for all the models
presented in Table\,2.  The intensity of the CIRB at 15$\,\mu$m for a
given population has been computed as,
$$
\mathrm{I}=\frac{1}{4\pi}\int{dL_{15}\int{dz \,\frac{dV}{dz}\, S_{\nu}(L_{15},z) 
\,\Phi(L_{15},z)}}
$$
where $S_{\nu}(L_{15},z)$ represents the observed flux of a source with an
intrinsic luminosity $L_{15}$ at redshift $z$.  The integration
has been performed for $\mathrm{log}\, L_{15}(z$=0)=[42,47] up 
to $z$=4. For type-2 sources the same redshift cutoff as for type-1 
($z_{cut}$=2.0) has  been assumed. Column X in Table 2 summarises the 
expected background light for each model  and the associated errors 
given by the 1$\sigma$ dispersion on the LF parameters as described in \S3.2. 

The integrated light from type-1 AGN galaxies provides a contribution
of (4.2--12.1)$\times 10^{-11}\;\mathrm{W m^{-2} sr^{-1}}$ (in units of
$\nu I_{\nu}$) at 15$\,\mu$m.  This value corresponds to (1.6--4.5)$\%$ of the
15$\,\mu$m background as reported by Elbaz et al. (2002) and confirmed by
Metcalfe et al. (2003; 2.7 $\times 10^{-9}\;\mathrm{W m^{-2} sr^{-1}}$) 
using deep lensed observations with ISOCAM. Our predictions are in agreement
with what previously estimated by M02 ($5.7\times 10^{-11}\;\mathrm{W m^{-2}
sr^{-1}}$).

The contribution of type-2 sources is found to be (5.5--14.6)$\times
10^{-11}\,\mathrm{W m^{-2} sr^{-1}}$, depending significantly on the
assumed SED and the parameterisation for the evolution of the faint
slope [$\alpha(z)$], and accounts for (2.0--5.4)\% of the CIRB at
15$\,\mu$m. We note here that, even if the space density of
the obscured sources, up to redshift 0.7 (where type-2 AGN are
detected), is a factor 2--6 higher than that of the un-obscured ones
(in the luminosity range $L_{15}\sim$[43.5,45.0], around L$^\ast$,
where most of the CIRB is produced), their contributions to the
CIRB at 15$\,\mu$m are very similar. The main reason is due to the
fact that type-2 AGN have a strong $k$-correction imposed by the
mid-IR SED.  Moreover, the effect is increased by a slightly lower
evolution rate in comparison to the type-1 AGN, and a significant
decline of the faint space density at higher $z$.

The total estimated contribution of AGN represents  $\sim$4--10\% of 
the total light observed in the mid-IR. This fraction is about half of 
the one ascribed to the AGN by M02.  M02 study was  limited to
type-1 AGN, and assumptions based on the unified model and the local 
ratio of type-2 to type-1 ($\sim$4; Maiolino \& Rieke 1995) were made 
to derive their contribution.  The smaller ratio of type-2 to type-1 
found in our fields can explain the observed difference in our estimates.

Mid-IR studies based on {\it IRAS} and {\it ISO} observations have
indirectly estimated a contribution of AGN to the CIRB not larger than
5--10$\%$ (e.g. Franceschini et al. 2001; Xu et al 2001, 2003). This is
the maximum room 'left' in their models by the strongly evolving
starburst population. The total contribution was in any case uncertain
since mid-IR selected type-2 AGN and starburst galaxies were treated
as a single population. Another estimate of this contribution comes
from the X--ray band (0.5--10 keV), which offers a better wavelength
regime where to select and identify obscured sources (unless the
sources are Compton-thick, $N_H \! > \! 10^{25}$\,cm$^{-2}$).
A cross-correlation of X--ray and IR sources detected by deep observations in 
the {\it Chandra} Deep field North (CDFN; Brandt et al. 2001) and in the 
Lockman Hole (Hasinger et al. 2001) allowed Fadda et al. (2002) to estimate 
the maximum fraction of the CIRB produced by AGN. This fraction is found to 
be of (17$\pm$6)\%, a factor 2--4 higher than our predictions. 

On the basis of the hard X--ray determination of the LF and
semi-empirical SEDs (linking the X--ray to the Infrared),
Silva et al. (2004) have derived the contribution of AGN and their
{\it host} galaxy to the CIRB. For type-1 AGN, the agreement between
their results ($\sim 7 \times \! 10^{-11}\;\mathrm{W m^{-2} sr^{-1}}$) and
the ones presented here is very good.

There is instead a significant difference  for type-2 sources. The predictions
from Silva et al.  (2004; $3 \times 10^{-10}\;\mathrm{W m^{-2} sr^{-1}}$) are,
in the most favourable case (model ``E''), at least a factor of 2 higher
than ours and closer to the numbers provided by Fadda et al. (2002). The higher 
efficiency of the hard X--ray observations to select obscured AGN can 
explain the discrepancy between our results and the ones presented by 
Fadda et al. (2002) and Silva et al. (2004).

The conclusion drawn from the comparison with the above mentioned
studies is that, as discussed in \S 3.4, it is likely that a significant 
fraction of type-2 AGN is hidden within the rest of the mid-IR selected population, 
the normal and starburst galaxies. 

\section{Mid-IR counts predictions for {\it Spitzer}}

\begin{figure*}[!t]
\resizebox{\hsize}{!}{\includegraphics{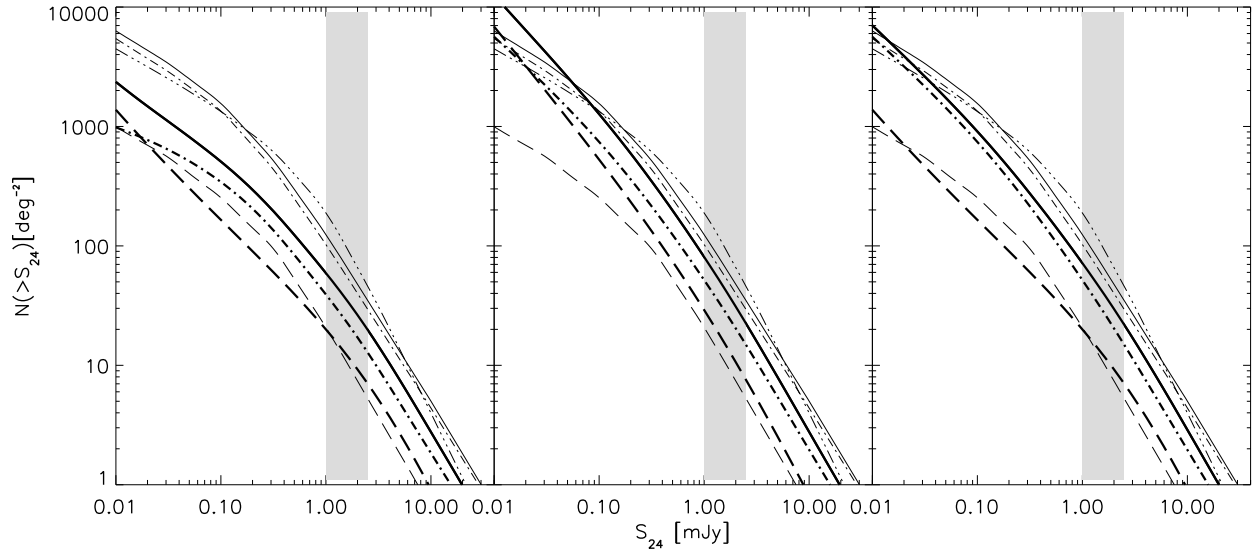}}
\caption{\scriptsize{The expected {\it Spitzer} MIPS integral counts 
for AGN at 24$\,\mu$m based on our best fit models (thick-{\it dashed} 
for type-1, thick-{\it dot-dashed} for type-2 and thick-{\it continuous} 
line for the total). The panels represent different combinations of type-1 
and type-2 models from Table 2. {\it Left} Models ``B'' and ``F'' for 
type-1 and type-2 respectively. {\it Central} Models ``A'' and ``E''. 
{\it Right} Models ``B'' and ``E''. For comparison, the expectations from 
the different AGN classes derived by Silva et al. (2004) are also shown 
(Type-1: thin $dashed$ line; Type-2: thin $dot$--$dashed$; Total: 
thin $continuous$). The total AGN contribution by Xu et al. (2003) 
is represented by a thin $triple$--$dot$--$dashed$ line. The light-grey 
shaded areas represent the expected flux range at 24$\mu$m of a source with
a 15$\mu$m flux of 1\,mJy given the assumed type-1 and type-2 SED.}}
\label{fig:Expected_24mu_counts}
\end{figure*}

The {\it Spitzer} space telescope is currently performing several deep mid- and
far-infrared observations in selected areas of the sky. Following the results
presented in this paper we have estimated the expected number of AGN sources as
a function of the flux in the {\it Spitzer} mid-IR band. The derived integral
counts for the MIPS instrument at 24$\,\mu$m are shown in Figure\,10. The
results are given in three panels that correspond to three different
combinations of type-1 (thick {\it  dashed}) and type-2 (thick {\it dot-dashed})
models (Table 2). The total AGN contribution (type-1\,+\,type-2) is given by a
thick {\it  continuous} line. Our estimates are compared to the total
contribution of AGN from the models of Xu et al. (2003) and Silva et
al. (2004). These models are plotted as a thin {\it triple-dot-dashed} (Xu et
al. 2003) and thin {\it continuous} (Silva et al. 2004) lines.  In the case of
the Silva et al. (2004) model the individual contributions from type-1 (thin
{\it dashed} line) and type-2 (thin {\it dotted-dashed} line) AGN are also
shown. Best fit solutions with a redshift dependence on the faint slope
(fits ``B'' and ``F'' for type-1 and type-2 respectively) are plotted in the
{\it left} panel. The {\it central} panel presents the expectations assuming the
best fits without any evolution for the faint slope of the LF (PLE models, fits
``A'' and ``E'').  Finally, the combination of type-1 and type-2 models that
better fits the expectations from Silva et al. 2004 (fits ``B'' for type-1 and
``E'' for type-2) is given in the {\it right} panel.

At a 24$\,\mu$m flux of 0.01 mJy the total AGN counts given by our
best fit solutions underestimate the predictions by Silva et al. (2004)
and Xu et al. (2003) by a factor of $\sim$2--3 (Fig.\,10, {\it left}).
With these models we predict to find $\sim$1200$^{+420}_{-300}$ 
type-1 and $\sim$1000$^{+350}_{-250}$ type-2 optically classified 
AGN per sq. degree down to a {\it Spitzer} flux limit of 
$S_{24\mu m}=0.01$ mJy.
If compared to the individual contributions of type-1 and type-2 AGN
given by the Silva et al. (2004) model, we note that the large
disagreement is mainly due to the low number of type-2 AGN expected in
our models (a factor $\sim$5 lower) and caused by the rapid flattening
with $z$ of the faint LF slope. The large effect of the LF faint slope
decline in the counts is evident if compared to the PLE models
predictions plotted in the {\it central} panel. The PLE models
overestimate the 24\,$\mu$m counts by both Xu et al. (2003)
and Silva et al. (2004) below 0.06 mJy.  A combination of the model
``B'' (variable LF faint slope for type-1 AGN) and model ``E'' (fixed
faint slope model for type-2 AGN) produces the best approximation to
the counts derived by the Silva et al. (2004) model based on the AGN
hard X--ray LF (Fig.\,10, {\it right}). If this is the case, then $\sim$1200 
type-1 and $\sim$5600 type-2 AGN per sq. degree are expected brighter 
than 0.01 mJy at 24\,$\mu$m. 

We want to note here that: i) although the expected {\it Spitzer}/24\,$\mu$m
integral counts derived by our different models differ by factors as large as
$\sim$6--10, they are very similar down to the flux limits of the ELAIS-South
Surveys (shaded area in Fig.\,10), responsible for the majority of the sources
observed at high redshift in our sample, ii) even if some of our models produce
counts comparable to the ones derived from X--ray LFs at $S_{24\mu
\mathrm{m}}=0.01$ mJy (Fig.\,10, {\it central} and {\it right} panel), they
are always lower at brighter fluxes. This second point explains the observed
difference (by a factor of $\sim$3) between our models and Silva et al. (2004)
in the contribution of AGN to the CIRB.

\section{Conclusions}

Combining {\it ISO}/15$\,\mu$m observations in ELAIS fields, HDF-N, HDF-S and
the CFRS with the local IR population detected by {\it IRAS} at 12$\,\mu$m, we
have derived the evolution for the mid-IR selected  and optically classified
AGN. While a similar study has already been done for QSOs+Seyfert-1 sources in
the past (Matute et al. 2002), we have presented here for the first time the
rate of evolution shown by the obscured, type-2 AGN. Our results are briefly
summarised as:
\begin{itemize}
\item[-]{Type-1 AGN evolve following a double power law LF and a rate of
	evolution of the form (1+$z$)$^{k_L}$ with $k_L\sim$2.9. The data are
	consistent, within the errors, with a PLE model. However,  a flattening
	at high redshift of the faint luminosity end of the luminosity function 
	is marginally favoured by the data.}
      \item[-]{Type-2 sources evolve with a slightly lower rate than
        type-1, with $k_L$ ranging from $\sim$1.8 to $\sim$2.6 depending
        strongly on the assumed SED used to compute the K-correction.
        The best fit solution favours a luminosity dependent luminosity 
	evolution (LDLE) model, while the PLE model is statistically 
	acceptable. However, integral counts reported for the deep 
	15$\,\mu$m observations in the HDF-N, -S and CFRS suggest an 
	evolutionary scenario closer to the PLE model. We expect a significant 
	number of type-2 AGN to be hidden within the optically classified 
	normal and starburst galaxies.}
      \item[-]{In the volume of the universe where both types of
        sources (type-1 \& type-2) are observed ($z$=[0,0.7]) the
        ratio of type-2 to type-1 is $\sim$2--6.  This value is
        in agreement, within the errors, with the one measured
        from the local optically selected samples, where a ratio
        close to 4 is found (Maiolino \& Rieke 1995).}
      \item[-]{The total contribution of AGN to the CIRB at 15$\,\mu$m
	is of the order of $\sim$4--10$\%$ of the total background light
        measured by Metcalfe et al. (2003), divided approximately equally
	between the two classes (type-1 \& type-2). The contribution of
        type-1 sources is in good agreement with the results presented in
	previous studies (Matute et al. 2002). Comparison with results from
	X--ray selected samples (e.g. Fadda et al. 2002, Silva et al. 2004)
	shows that a significant fraction of obscured type-2 sources are missed
	in our sample. We argued that the optical spectral classification of the
	mid-IR sample, and not the mid-IR selection, is the principal
	responsible for the missed fraction.}
      \item[-]{Estimates for the number of mid-IR selected and
          optically classified AGN, expected from {\it Spitzer} at
          24$\,\mu$m, are given according to the best model results.
          We expect $\sim$1200 type-1 and $\sim$1000 type-2 optically 
          classified AGN per sq. degree at a flux limit of $S_{24\mu m}=0.01$ mJy. 
          Our 24\,$\mu$m counts derived for type-1 sources agree very well 
          with previous works. At a flux limit of $S_{24\mu m}=0.01$ mJy the
          expected counts for the obscured population from our best fit models 
          are a factor $\sim$5 lower than the expectations from models based on
          X--ray LFs (e.g Silva et al. 2004).}
\end{itemize}

The results presented here for type-1 sources are quite robust since a high 
completeness is expected for these sources. Unlike type-1, type-2 AGN show a 
larger spread in their mid-IR and optical  properties and a significant fraction 
can be misclassified as starburst or normal galaxies. As a consequence, the 
results for type-2 AGN derived here can only be considered as a lower limit to
their true density and a first approximation to the evolution of these sources
in the mid-IR. The true mid-IR space density of obscured sources has to be
determined combining their mid-IR properties and the optical classification with
the information available at other wavelengths, especially in the X--rays.

\begin{acknowledgements}

The authors are grateful to the referee for helpful comments and constructive
criticism improving the manuscript.  This paper is based on observations
collected at the European Southern Observatory, Chile (ESO No. 57.A-0752,
58.B-0511, 59.B-0423, 61.B-0146, 62.P-0457, 67.A-0092(A), 68.A-0259(A),
69.A-0538(A) and 70.A-0362(A).  This research has been partially supported by
ASI, INAF and MIUR grants.  I.\,M. acknowledges a Ph.D.  grant from CNAA/INAF.

\end{acknowledgements}

\end{document}